# Scintillation Reduction for Laser Beams Propagating Through Turbulent Atmosphere

## G.P. Berman[1], V.N. Gorshkov[1,2,3], S.V. Torous[2]

[1]Theoretical Division, T-4 & CNLS MS B213, Los Alamos National Laboratory, Los Alamos, New Mexico 87545
[2]National Technical University of Ukraine "KPI," 37 Peremogy Avenue, Building 7, Kiev-56, 03056 Ukraine
[3]The Institute of Physics, National Academy of Sciences of Ukraine, Nauky Ave. 46, Kiev 680028, Ukraine

E-mail: **gpb@lanl.gov**

**Abstract**

We numerically examine the spatial evolution of the structure of coherent and partially coherent laser beams, including the optical vortices, propagating in turbulent atmospheres. The influence of beam fragmentation and wandering relative to the axis of propagation (*z*-axis) on the value of the scintillation index (SI) of the signal at the detector is analyzed. These studies were performed for different dimensions of the detector, distances of propagation, and strengths of the atmospheric turbulence. Methods for significantly reducing the scintillation index are described. These methods utilize averaging of the signal at the detector over a set of partially coherent beams (PCBs). It is demonstrated that the most effective approach is using a set of PCBs with definite initial directions of propagation relative to the *z*-axis. This approach results in a significant compensation of the beam wandering which in many cases is the main contributor to the SI. A novel method is to generate the PCBs by combining two laser beams - Gaussian and vortex beams, with different frequencies (the difference between these two frequencies being significantly smaller than the frequencies themselves). In this case, the effective suppression of the SI does not require high-frequency modulators. This result is important for achieving gigabit data-rates in long-distance laser communication through turbulent atmospheres.

**Introduction**

Laser beam scintillations caused by atmospheric turbulence are a major concern for gigabit data-rates and long-distance optical communications [1,2]. These scintillations are characterized by the scintillation index (SI),

$$\sigma^2(z) = \left\langle \left(I^l(z)\right)^2 \right\rangle_l \Big/ \left\langle I^l(z) \right\rangle_l^2 - 1. \qquad (1)$$

Here, $I^l(z) = \dfrac{1}{S}\int_S I^l(x,y,z)dxdy$ is the signal intensity at the detector with area of $S$; the superscript, *l*, indicates a particular state of the atmospheric turbulence; and the subscript, *l*, indicates an average over many such states. The optical strength of turbulence is described by index of turbulence, $C_n^2$, which is associated [3] with the structure function, $D(r)$, for fluctuations of the refractive index, $n'(\vec{r})$:

$$D(r) = \left\langle \left[ n'(\vec{r}' + \vec{r}) - n'(\vec{r}') \right]^2 \right\rangle_l = C_n^2 r^{2/3}.$$

One of the contributors to the SI is beam fragmentation, which is the decay of the initial laser beam into many spatially separated beams. (See Fig. 1.) There is a probability that these separated beams cannot be detected if the detector size is smaller than the characteristic distance between the beams. The coherent beam is initially oriented along the z-axis. Depending on $l$, this beam deviates in the x-y-plane. This deviation is defined by the beam wandering (see [4] and references therein):

$$\vec{r}_w^l(z) = \int \vec{r} I^l(x,y,z) dx dy \Big/ \int I^l(x,y,z) dx dy . \qquad (2)$$

The wandering of the beam significantly contributes to the scintillations, $\sigma^2$, of intensity, $I^l$, on the detector [5].

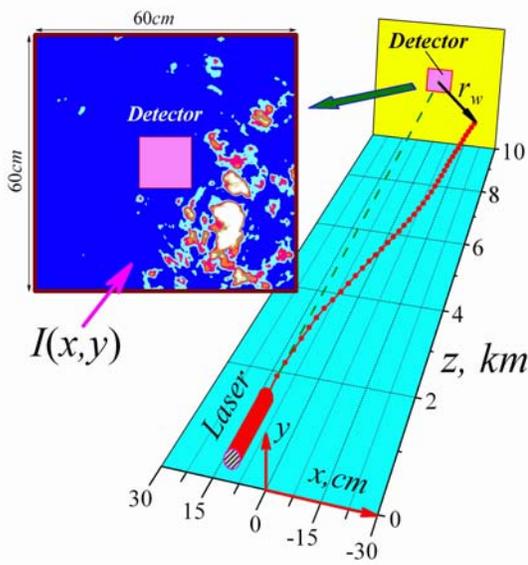

**Figure 1.** Numerical simulation of the propagation of an initially coherent laser beam through a single realization of a turbulent atmosphere ($\lambda = 1.55 \mu m$; initial laser–beam radius, $r_0 = 2 cm$; $C_n^2 = 2.5 \times 10^{-14} m^{-2/3}$). Note the fragmentation at the detector and the beam wandering.

It is known that a partially coherent beam (PCB) in combination with a slow time response detector leads to a significant reduction of the SI. (See [6-11] and references therein.) The idea of the method is the following. For a given state of the atmosphere, $l$, a coherent laser beam passes through the phase modulator (PM), which randomly changes the phase of the beam, $\varphi_m(x,y)$ $(m = 1, 2, ..., M)$ in the aperture plane ($m$ is the state of the PM with a corresponding value of the intensity, $I_m^l$, on the detector; $M$ is the number of realizations integrated by the detector). The average over the PM states signal at the detector; $\hat{I}^l = \frac{1}{M} \sum_{m=1}^M I_m^l$, can become quite stable—almost independent of the atmospheric realizations, $l$. Namely, $\hat{\sigma}^2 \ll \sigma^2$, where $\hat{\sigma}^2$ is the SI calculated by averaging $\hat{I}^l$ over $l$.

The effectiveness of this approach was experimentally demonstrated in [10,12] for a simplified model in which the atmospheric turbulence was simulated by a spatial light modulator (atmospheric modulator, AM). A proper choice of PM allowed the authors of [12] to reduce the SI by a factor of 16 (for $M = 10$) This SI reduction is caused by the formation of significantly smaller-scale speckle structures after the PCB passes the AM, compared with the similar situation for a coherent beam. However, the optimistic theoretical predictions were done in [6,7,8] only for the case of strong turbulence and long propagation distances ($z > L_{thres}$, where $L_{thres}$ is the threshold used in the theory, which based on an asymptotic method). In the region where $\sigma^2(z)$ approaches its maximum values, the theory [6,7,8] is not applicable ($z < L_{thres}$).

In this paper we analyze the case of moderate laser propagation distances, $z < L_{thres}$, and we discuss how beam fragmentation and wandering affect the SI, $\sigma^2$. The main objective is to design algorithms for the PM that allow us to significantly reduce SI for a relatively small number of realizations, $M$. The latter is important for practical implementations of long-distance optical communications. In the process of designing of these algorithms, we have examined the propagation of optical vortices (OV) [13] in turbulent atmospheres. As described in detail below, we recommend the use of a PCB that combines an OV with a Gaussian beam (GB). If the frequencies of these types of beams differ by $\delta\omega$, then for a stationary distribution of the phase mask, $\varphi_0(x,y)$, a significant reduction of the SI can be achieved for the time-averaged scintillations of intensity, $\hat{I}^l = \frac{1}{T}\int_0^T I^l(t)dt$, where $T = 2\pi/\delta\omega$.

## 1. Spatial evolution of a coherent laser beam propagating through a turbulent atmosphere

Here we will investigate in the paraxial approximation the spatial evolution of a stationary linearly polarized light beam propagating along the $z$-axis with vector potential $\mathbf{A} = \mathbf{e}U(x,y,z)\exp[i(\omega t - kz)]$. (Here $\mathbf{e}$ is the unit vector in the direction of light polarization; $\omega$ is its frequency; and $k = 2\pi/\lambda$ is its wave vector.) The values of the magnetic and electric fields, $\mathbf{B}$ and $\mathbf{E}$, are given by $\mathbf{B} = \nabla \times \mathbf{A}$ and $\mathbf{E} = -\frac{1}{c}\frac{\partial \mathbf{A}}{\partial t}$. The light intensity is $I(x,y,z) \sim |U(x,y,z)|^2$.

A turbulent atmosphere can be modelled by a system of random phase screens [14-17]. After passing each of these screens, the beam experiences phase distortions in the $(x,y)$-plane. The distance between neighboring screens is $50 m$. Between the screens, the complex amplitude, $U(x,y,z)$, satisfies the Leontovich parabolic equation,

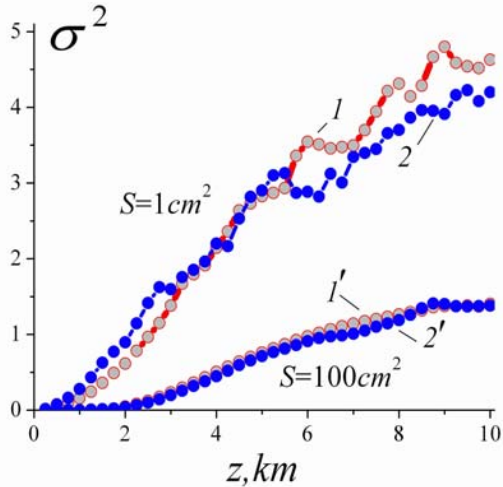

**Figure 2.** $\sigma^2(z)$ for a Gaussian beam with radius of $r_0 = 2cm$ (curves $1,1'$) and $r_0 = 5cm$ (curves $2,2'$). Averaging was done over 5000 random atmospheric states, $l = 1,2,...,L_{atm} = 5000$.

$$\frac{\partial U}{\partial \tilde{z}} = \frac{1}{4i}\left[\frac{\partial^2 U}{\partial \tilde{x}^2} + \frac{\partial^2 U}{\partial \tilde{y}^2}\right]. \tag{3}$$

We use the dimensionless coordinates, $\tilde{z} = z/z_R$, $\tilde{x} = x/r_0$, $\tilde{y} = y/r_0$, where $z_R = \frac{\pi r_0^2}{\lambda}$ is the Rayleigh length; $r_0$ is the beam radius; and $\lambda = 1.55 \mu m$ is the wavelength. The optical field in the plane $z = 0$ has the Gaussian distribution, $U(x, y, z = 0) = U_0 \exp(-r^2/r_0^2)$.

Equation (3) is solved numerically using a difference method, the Pismen-Reckford scheme being employed [18]. Our finite-difference grid has the following uniform spacings:

$$-\frac{L}{2} \leq x \leq \frac{L}{2}, \quad -\frac{L}{2} \leq y \leq \frac{L}{2}, \quad 0 \leq z \leq 10km,$$

$$L = N\Delta x = 3.6m, \quad N = 7200,$$

$$\Delta x = \Delta y = 0.5mm \ll l_0 = 6.5mm, \quad \Delta z = 1m \sim z_R \times 10^{-3},$$

where $l_0$ is the inner scale of turbulence.

In our numerical simulations, $\Delta x$ and $\Delta y$ are sufficiently small, so that the amplitude, $U(x, y, z)$, (not the intensity $|U(x, y, z)|^2$) varies slowly on this scale.

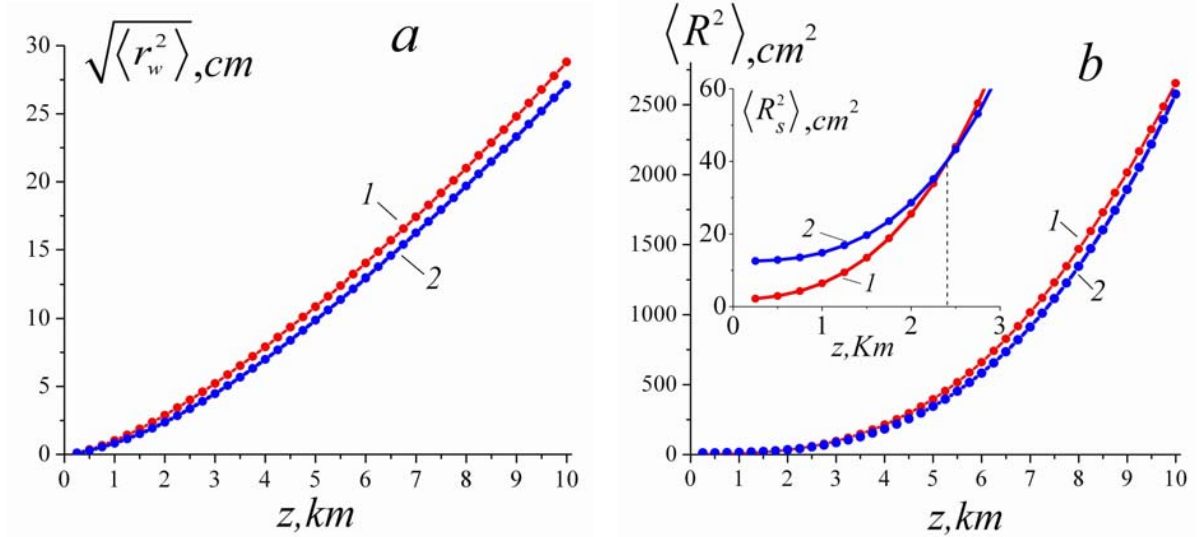

**Figure 3**. Spatial evolution of the geometric characteristics of laser beams. At $z = 0$ $R = r_0/\sqrt{2}$. Curves *1* and *2* have $r_0 = 2cm$ and $5cm$, correspondingly. The diffraction spreading of the first beam is larger, because its Rayleigh length $z_R(r_0 = 2cm) \approx 811m$, whereas $z_R(r_0 = 5cm) \approx 5067m$. (See insert in Fig. 3*b*.)

At the boundary of the computational space, we set $U(\tilde{x}, \tilde{y}, \tilde{z}) = 0$. We do not use the Fourier expansion (FFT subroutine in the IMSL library) for solving Equation (3) because this method has significant errors for complicated fields and a large number of nodes. The signals, $I^l$, are calculated for two detector integration areas ($S = 1\times 1 cm^2$ and $S = 10\times 10 cm^2$) both having their centers on the $z$-axis. Most of numerical simulations presented in this paper were done for a moderate level of turbulence, $C_n^2 = 2.5\times 10^{-14} m^{-2/3}$. In Section 6 the case of stronger turbulence, $C_n^2 = 10^{-13} m^{-2/3}$, is investigated.

The characteristic features of propagation of a coherent laser beam in a turbulent atmosphere are presented in Fig. 2. In particular, for the small detector, $S = 1 cm^2$, one can see that $\sigma^2(z; r_0 = 5cm) > \sigma^2(z; r_0 = 2cm)$ at relatively small propagation distances to the detector, $z < 4km$. For $z > 5km$, the opposite dependence is observed: $\sigma^2(z; r_0 = 5cm) < \sigma^2(z; r_0 = 2cm)$. The physical mechnisms responsible for this dependence of the SI on propagation distance, $z$, will be discussed below. Now we will discuss the geometric characteristics of laser beams, which are presented in Fig. 3 and where

$$\langle R^2(z) \rangle = \left\langle \int r^2 I(x,y,z) dx dy \Big/ \int I(x,y,z) dx dy \right\rangle_l. \qquad (4)$$

The root-mean-square radius $\sqrt{\langle R^2 \rangle}$ characterizes the size of the optical field.

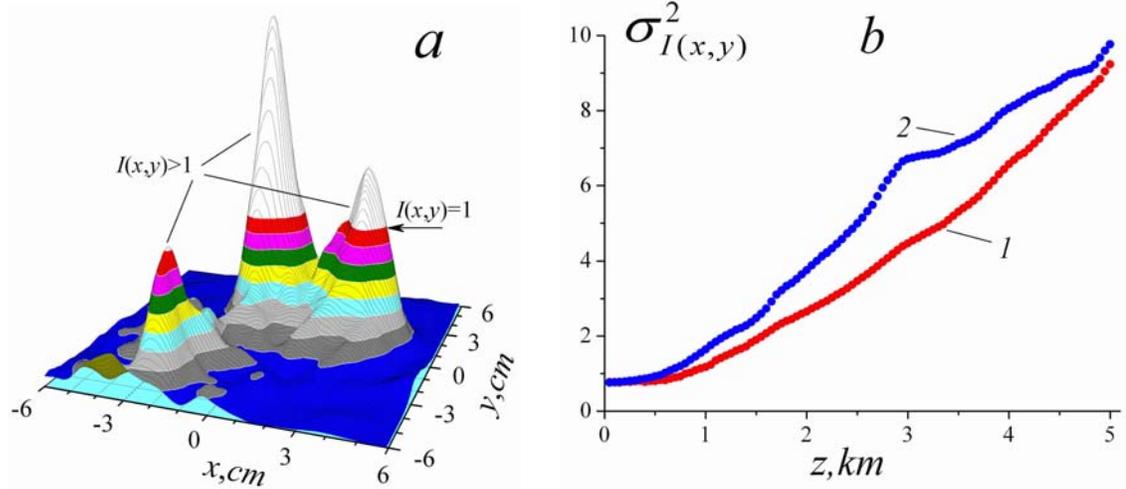

**Figure 4.** $a$ – Distribution of beam intensity in the plane, $z = 2km$, for a single atmospheric realization; $r_0 = 5cm$, $\sigma^2_{I(x,y)} = 4.33$ (for $r_0 = 2cm$ and for this atmospheric realization, $\sigma^2_{I(x,y)} = 2.84$). $b$ – Dependence of $\sigma^2_{I(x,y)}$ on distance, which characterizes the inhomogeneous distribution of intensity in the cross section of the beam. 1 and 2 – indicate $r_0 = 2cm$ and $5cm$, respectively.

A qualitative explanation of why the narrower beam has a larger deviation, $r_w$, than the broader beam (Fig. 3a) for a fixed atmospheric state, $l$, is the following. The deviation of the beam from its initial $z$-direction is mainly due to inhomogeneities of the index of refraction, $n'$, with characteristic scales, $\Lambda$, larger than the effective beam radius, $R_s(z)$, which characterizes the size of the optical field in the coordinate system centered at $\vec{r}_w(z)$:

$$R_s^2(z) = R^2(z) - r_w^2(z).  \qquad (5)$$

The small-scale inhomogeneities with $\Lambda << R_s$ only scatter the beam. Thus, the smaller deviation of the broader beam is caused by its narrower "effective zone", $q < q_W \equiv 2\pi / R_s$, in the spectrum of turbulent fluctuations of refractive index [14-17], $\Phi_n(q)$, which determines the magnitude of deviation of the beam:

$$\Phi_n(q) = 0.033 C_n^2 q^{-11/3} \exp(-q^2 / q_m^2), \qquad (6)$$

where $q_m = 5.92 / l_0$, $q = 2\pi / \Lambda$ is the wave vector corresponding to the inhomogeneities of $n'$ with wavelength, $\Lambda$.

The difference in wandering, $\langle r_w^2 \rangle_{r_0=2cm}$, $\langle r_w^2 \rangle_{r_0=5cm}$, observed in Fig. 3$a$ is small. The reason is that $\langle R^2 \rangle$ (and, consequently, $\langle R_s^2 \rangle$ and $q_W$) are practically independent of the initial beam radius, at large distances (Fig. 3b) because the beam spreading due to the turbulent atmosphere is significantly larger than the spreading due to beam diffraction. These results are not only in qualitative but also in quantitative agreement with the analytic theory [4,6], according to which,

$$\langle R^2 \rangle = r_0^2 (1 + z^2 / z_R^2)/2 + 2.2 C_n^2 z^3 l_0^{-1/3}. \qquad (7)$$

For the cases under consideration, the second term in (7) begins to exceed the first term (which describes diffraction spreading) for $z > 1.65 km$. In this same region, one can observe that the value of $\sigma^2(z; r_0 = 5cm)$ exceeds $\sigma^2(z; r_0 = 2cm)$ for detectors with $S = 1 cm^2$. (See Fig. 2). The question is: What is the cause of this, because at first sight the opposite inequality should have been satisfied: $\sigma^2(z; r_0 = 5cm) < \sigma^2(z; r_0 = 2cm)$?

The answer is the following. The region of wave vectors, $q < q_W$ (or $\Lambda > R_s$) defines the region of beam wandering. The small-scale inhomogeneities, $\Lambda < R_s$, $q > q_W$, of the refraction index determine the redistribution of the intensity over the beam crosssection, *i.e.* its fragmentation. For relatively small distances, $z$, $q_{W, r_0=5cm} < q_{W, r_0=2cm}$. The fragmentation of the broader beam will be more significant, and the SI will be larger. This statement is illustrated in Fig. 4. In our simulations, the amplitude of the initial field distribution, $U_0$, was equal to unity. Consequently, $I_{max}(z=0) = I(x=0, y=0, z=0) = 1$. Fig. 4$a$ represents one of the variants of the laser beam fragmentation (with $r_0 = 5cm$), when the regions with $I(x, y) > 1$ were realized.

In order to compare the levels of fragmentation of the narrower and broader beams, we calculated for each of them the SI, $\sigma^2_{I(x,y)}(z)$. (See Fig. 4b.) The procedure of calculation was the following. For the atmospheric state, $l$, the value $\sigma^2_{I(x,y),l}$ was calculated for a cross section at a given $z$:

$$\sigma^2_{I(x,y),l} = \frac{\left\langle \left(I^l(x,y)\right)^2 \right\rangle_{x,y} - \left\langle I^l(x,y) \right\rangle^2_{x,y}}{\left\langle I^l(x,y) \right\rangle^2_{x,y}}. \qquad (8)$$

The averaging was performed over the circle in the $xy$-plane containing 95% of the total beam power. The center of this circle is located at $\vec{r}^{\,l}_w(z)$. Then, these results are averaged over all atmospheric realizations:

$$\sigma^2_{I(x,y)} = \frac{1}{L_{atm}} \sum_{l=1}^{L_{atm}} \sigma^2_{I(x,y),l}. \qquad (9)$$

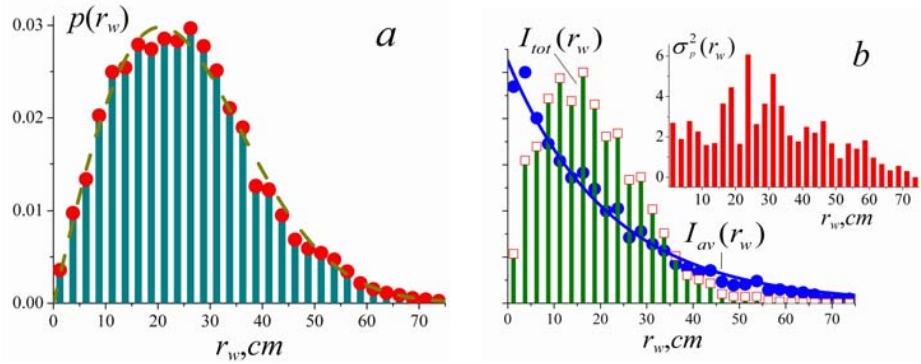

**Figure 5.** $a$ – The distribution function, $p(r_w)$, for a coherent beam ($r_0 = 2cm$) propagating initially along the $z$-axis, at $z = D$. The dashed line indicates an approximation of this distribution: $p(r_w) \approx 2.4 \times 10^{-3} r_w \exp\left[-r_w^2/(2r_{w,\max}^2)\right]$, $r_{w,\max}^2 = 410 cm^2$. $b$ – Statistical characteristics of intensity at the detector ($S = 1cm^2$) as functions of beam wandering $r_w$; $r_0 = 2cm$, $z = D$. $\bullet - I_{av}(r_w)$-average intensity (over atmospheric states, $l$) for beams with wandering $r_w$; $I_{av}(r_w) \sim \exp(-r_w/21)$; $\square - I_{tot}(r_w)$ is the integral intensity of all beams with wandering $r_w$: $I_{tot}(r_w) \sim I_{av}(r_w) \times p(r_w)$. In the insert: SI, $\sigma^2_p(r_w)$, calculated for sets of intensities, $\{I^l(r_w)\}$.

Larger values of $\sigma^l_{I(x,y)}$ (see Fig. 4b) correspond to more significant fragmentations of the beam, and larger values of the SI. (See Fig. 2, curves $1,2$.) However, for $z > 5km$, $\sigma^2(z, r_0 = 5cm)$ is smaller than $\sigma^2(z, r_0 = 2cm)$. In this region, the dominant role is played by the beam wandering, which is smaller for the beam with the initial radius, $r_0 = 5cm$. (See. Fig. 3a.) When the area of detector increases, (see Fig. 2, curves $1', 2'$) differences in values of SI have a smaller influence. But the value $\sigma^2(z=10km)$ is large enough, which also is due to the beam wandering. Indeed, as follows from Fig. 3, at $z=10km$, $\sqrt{\langle r_w^2 \rangle} \sim 30cm$, the characteristic size of the beam is $\sqrt{\langle R_s^2 \rangle} = \left[\langle R^2 \rangle - \langle r_w^2 \rangle\right]^{1/2} \sim 40cm$. Thus, the beam wandering is comparable with its width.

Our analysis demonstrates that in order to develop a more effective approach for reducing the SI, one requires a more detailed study (in addition to the $\sigma^2(z)$ discussion presented above) of statistical characteristics of $I^l(z)$ for given values of $C_n^2$. We performed this analysis for $z = D = 10km$. (See Fig. 5.) As demonstrated in Fig. 5a, the more probable values of beam wandering correspond to the interval, $10cm < r_w < 35cm$. We now examine quantitatively how much this interval contributes to the value of the SI, $\sigma^2$.

To analyze this problem, we consider a set of signals, $I^l$, as function of $r_w$, and divide this set into partial sets, $\{I^l(r_w)\}_k$. For elements of these subsets a condition is satisfied, $(k-1)\delta < r_w < k\delta$ for $k$ between 1 and 30. (We have chosen $\delta = 2.5cm$). In each subset we define its average value, $I_{av}(r_w)$ and the sum of all elements in the subset, $I_{tot}(r_w)$ (see Fig. 5b), and the corresponding SI, $\sigma_p^2(r_w)$. (Insert in Fig. 5b.) One can see that the partial sets with $15cm < r_w < 35cm$ are more "noisy" (have large values of $\sigma_p^2(r_w)$- see the insert). As our simulations show, this range of subsets accounts for $\sim 50\%$ of a total intensity, $\sum_{r_w} I_{tot}(r_w)$, at the detector.

***Conclusion***. In order to significantly reduce the SI one needs to "neutralize" the wandering in a certain region, $\Delta r_w(z)$, which depends on $C_n^2$. This dependency can be obtained either by numerical simulations or by analytic approaches. In the case considered above, ($C_n^2 = 2.5 \times 10^{-14} cm^{-2/3}$, $z = 10km$) the important values of wandering are concentrated in the region: $15cm < r_w < 35cm$. Note, that neutralization of these wanderings must be realized for a broad range of the distances, $z$, because $\sqrt{\langle r_w^2 \rangle}$ depends approximately linearly on $z$. (See Fig. 3a.)

Below we consider the effectiveness of three different approaches to solve this problem by averaging the signal over $M$ PCB realizations for a given atmospheric state, $l$.

## 2. Propagation of a PCB with random phase modulation through a turbulent atmosphere

Here we present the results of our numerical simulations for PCBs which are realized by a PM using phase masks, $\varphi_m(x,y)$, generated by the algorithm described in our paper [12]. One realization of the PM is shown in Fig. 6 *a*. The coherence radius of the PCB, $r^{(c)}$, in the plane $z = 0$ could be defined by calculating the correlation functions,

$$S_x(\delta) = \frac{\langle \varphi_m(x,y) \times \varphi_m(x+\delta, y) \rangle}{\langle [\varphi_m(x,y)]^2 \rangle}, \quad S_y(\delta) = \frac{\langle \varphi_m(x,y) \times \varphi_m(x, y+\delta) \rangle}{\langle [\varphi_m(x,y)]^2 \rangle},$$

which are identical in our case. According to results presented in Fig. 6$b$, $r^{(c)} \approx 4mm$, so the ratio $\left(r^{(c)}/r_0\right)^2 \approx 1/25$. (Here and below, $r_0 = 2cm$.)

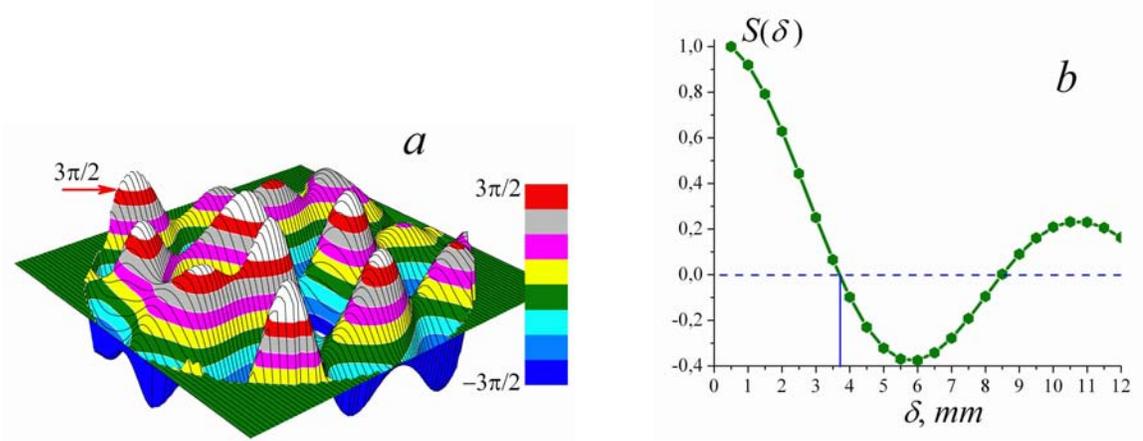

**Figure 6.** $a$ – One of the random phase distributions of the PM (only the interval $r \leq r_0$ is shown; $r_0 = 2cm$ is the radius of the Gaussian beam; $b$ - The correlation function of the PM.

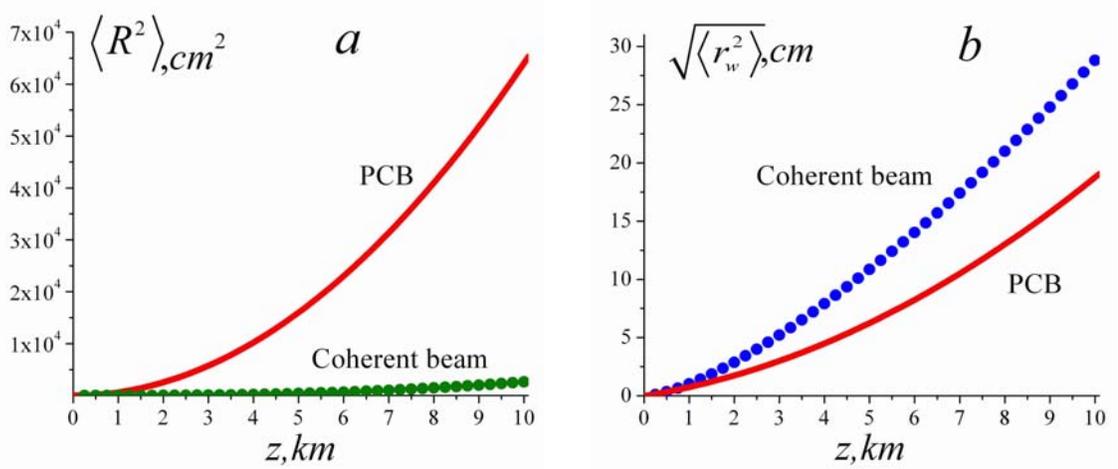

**Figure 7**. $a$ – The beam radius as a function of the propagation distance; $b$ – The wandering as a function of distance; $r_0 = 2cm$, $C_n^2 = 2.5 \times 10^{-14} m^{-2/3}$.

In this case, according to the theoretical results [7,8], the scintillations, $\hat{\sigma}^2$, of the average signal at the detector should be reduced by the factor: $\dfrac{\hat{\sigma}^2}{\sigma^2} \sim \left(r^{(c)}/r_0\right)^2$. Note that the theoretical results [7,8] are valid only for sufficiently large propagation distances, $z > L_{thres}$. Here we analyze the effectiveness of the PCB for relatively small distances where $\sigma^2(z)$ reaches its maximum for a Gaussian beam.

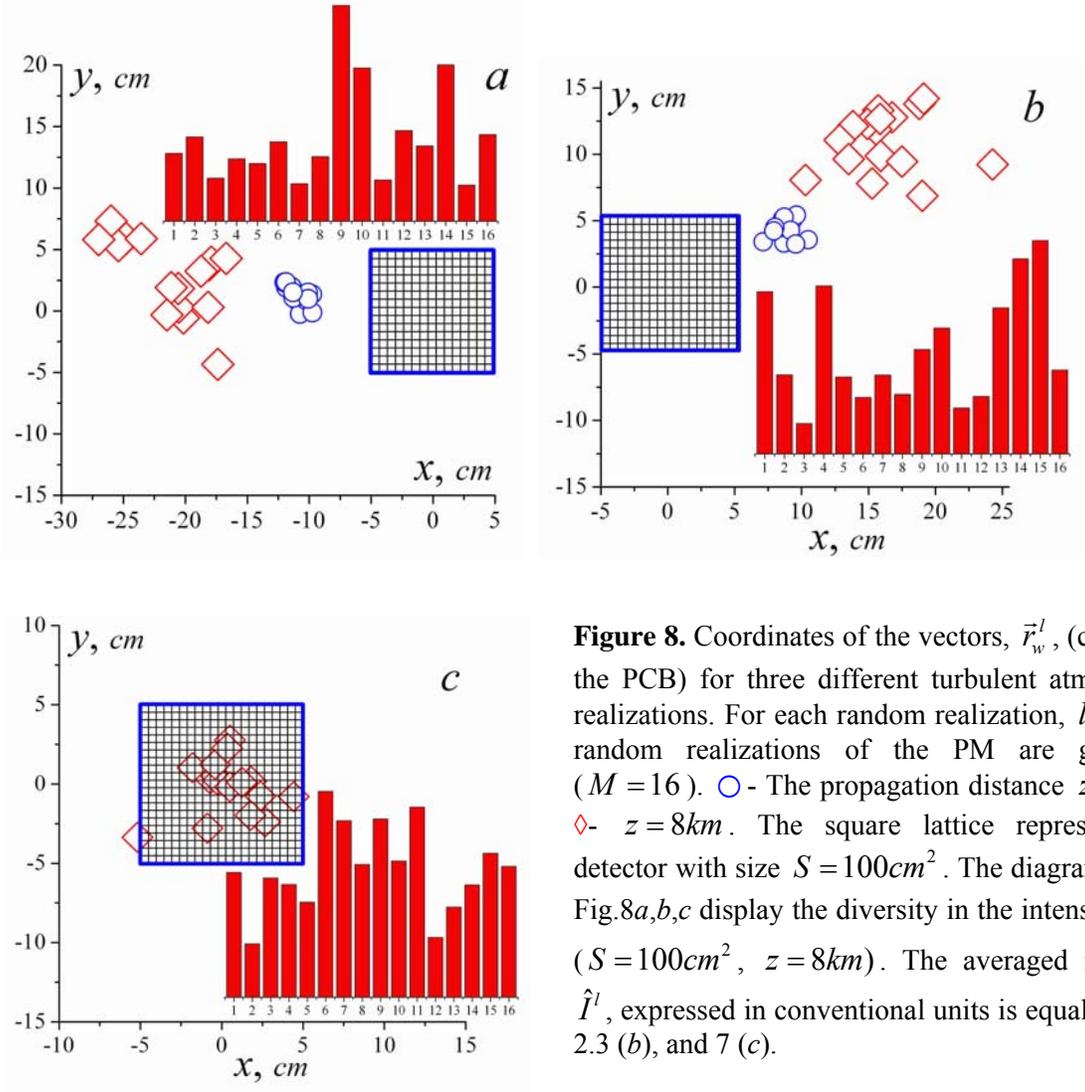

**Figure 8.** Coordinates of the vectors, $\vec{r}_w^l$, (centers of the PCB) for three different turbulent atmospheric realizations. For each random realization, $l$, sixteen random realizations of the PM are generated ($M = 16$). ○ - The propagation distance $z = 4km$, ◊ - $z = 8km$. The square lattice represents the detector with size $S = 100cm^2$. The diagrams in the Fig.8$a,b,c$ display the diversity in the intensities, $I_m^l$ ($S = 100cm^2$, $z = 8km$). The averaged intensity, $\hat{I}^l$, expressed in conventional units is equal to 1 ($a$); 2.3 ($b$), and 7 ($c$).

One can make the following assumptions about the possible mechanism of reduction of the SI. In the process of propagation, a PCB will spread significantly for a small coherence radius, $r^{(c)}$ (scattered by the phase mask). In this case, as we found, the effective region of wave vectors, $q < q_W \approx 2\pi / R_s$, in the spectrum of the refractive index, $n'(x, y, z)$ that is responsible for deviation of the beam, narrows. As a result, the value $\sqrt{\langle r_w^2 \rangle}$ falls compared to the deviation of the coherent beam. Averaging of the signal over $M$ realizations of the PM reduces the unwanted effects of beam fragmentation. Thus, a reduction of the SI, $\hat{\sigma}^2$, can be achieved. Our assumptions are confirmed by the data presented in Fig. 7a (the beam radius was significantly increased). In this case, the deviations of the beam were reduced. (See Fig. 7$b$.)

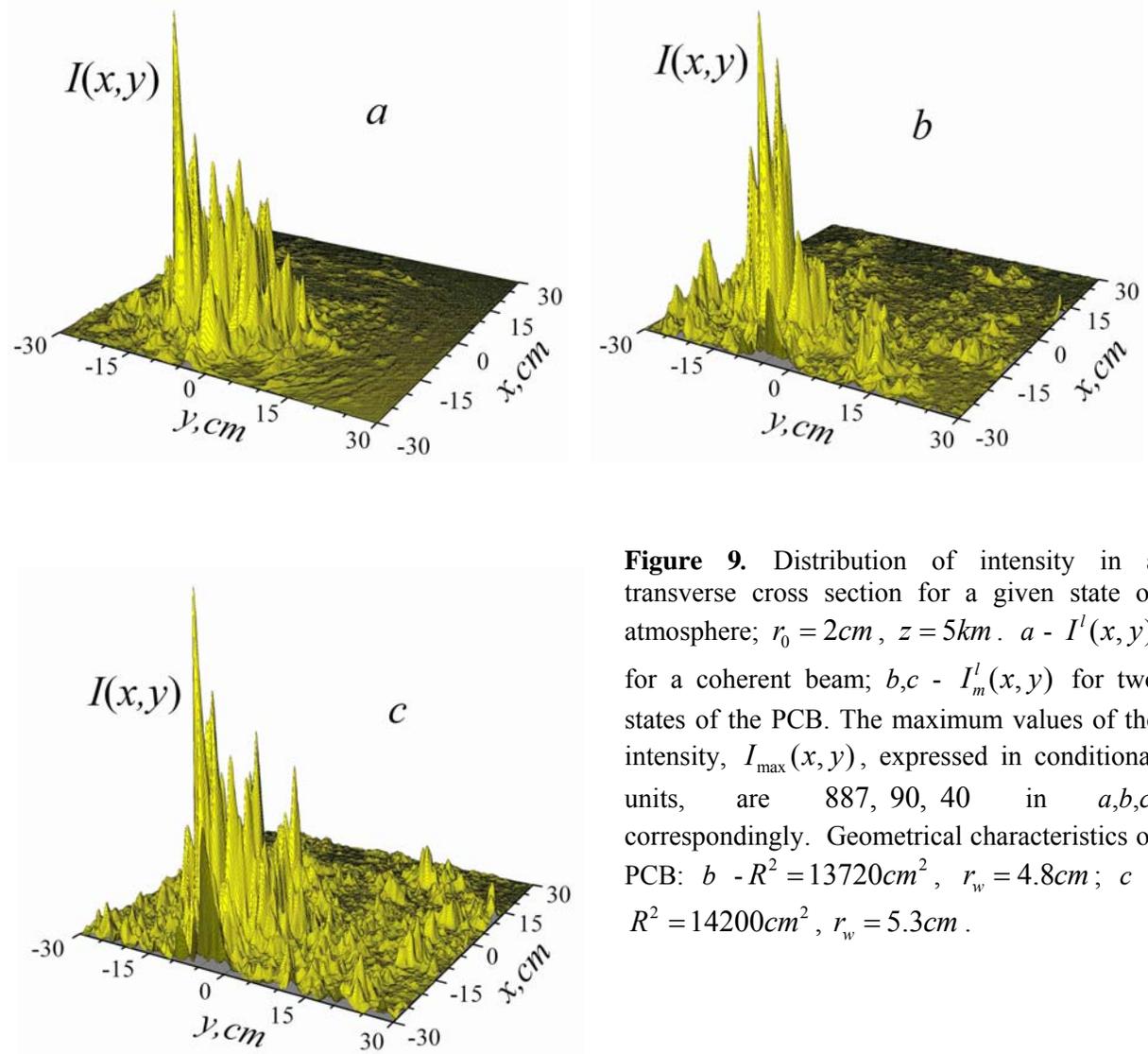

**Figure 9.** Distribution of intensity in a transverse cross section for a given state of atmosphere; $r_0 = 2cm$, $z = 5km$. $a$ - $I^l(x,y)$ for a coherent beam; $b,c$ - $I_m^l(x,y)$ for two states of the PCB. The maximum values of the intensity, $I_{max}(x,y)$, expressed in conditional units, are 887, 90, 40 in $a,b,c$, correspondingly. Geometrical characteristics of PCB: $b$ - $R^2 = 13720 cm^2$, $r_w = 4.8cm$; $c$ - $R^2 = 14200 cm^2$, $r_w = 5.3cm$.

Here an important point should be made. The effect of signal averaging over states, $m$, of the PCB is most effective when the intensities $I_m^l (m = 1, 2, ..., M)$ at the detector are statistically independent for a given atmospheric state, $l$. In this case, the SI for intensity, $\hat{I}^l$, is $M$ times smaller than the SI, $\hat{\sigma}_s^2$, calculated for the set $\{I_m^l\}$ [12] (in which $l$ and $m$ take all possible values).

But the statistical independence of these signals is not realized for a relatively small distance of propagation. (See Fig.8.) The trajectories of the centers of the beams, $\vec{r}_m^l(z)$, $m = 1, 2, ..., M$, are correlated (especially in the region $z \leq 5km$). See Fig. 8. Moreover, for a given atmospheric state, $l$, the distributions of intensity of a PCB with different states, $m$, often look visually similar and are similar to the distribution of intensity for a coherent beam. (See Fig. 9.) In each case, the examples of intensity distribution, $I(x,y)$, are presented in the region $-30cm < x < 30cm$, $-30cm < y < 30cm$.

As a result, the PCB is not really efficient in the considered case (see Fig. 10a, $\hat{\sigma}^2(z) \approx \sigma^2(z)$). The main reason is that the beam wandering, $\vec{r}_w(z)$, is not compensated in the process of beam propagation. The spreading for a PCB with small coherence radius, $r^{(c)} \ll r_0$, is mainly due to the influence of the PM (see Fig. 10b), and not due to turbulent atmosphere.

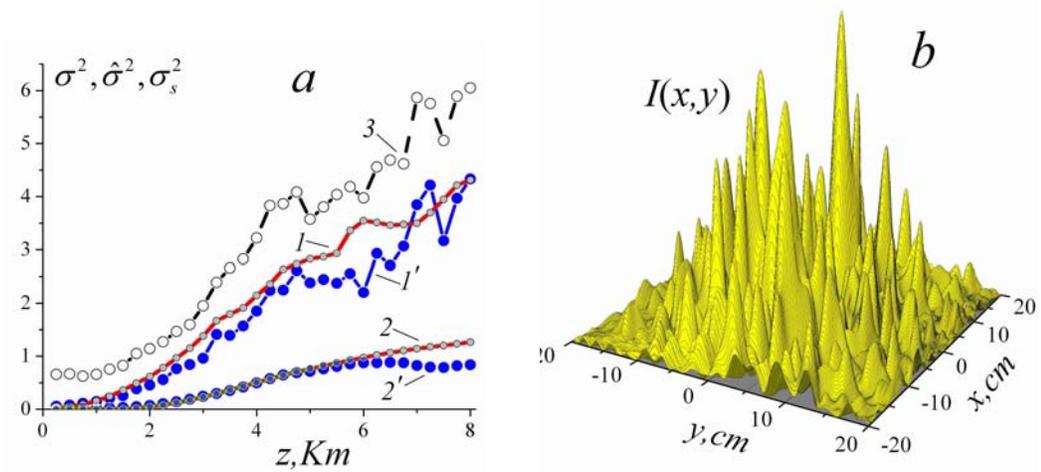

**Figure 10.** Results for a PCB with small initial coherence radius. *a* -Curves $1, 2 - \sigma^2(z)$ for coherent beam, $S = 1 cm^2$ and $100 cm^2$, correspondingly. $1', 2' - \hat{\sigma}^2(z)$ for the averaged intensity $\hat{I}^l(z) = \frac{1}{M}\sum_{m=1}^{M} I_m^l(z)$, $M = 16$, $L_{atm} = 250$. Curve 3 - $\sigma_s^2(z)$, calculated for a set, $\{I_m^l(z; S = 1cm^2)\}$, which contains $250 \times 16 = 4000$ elements. *b* – The characteristic distribution of the PCB intensity at a distance of $500 m$ in a homogeneous atmosphere (refraction coefficient $n = 1$). $r_w(z = 8km; n = 1) < 2cm$.

If the intensities at the detector, $I_m^l$, $m = 1, 2, ..., M$, were statistically independent, then the relation, $\hat{\sigma}^2 = \sigma_s^2/M$, would have been satisfied. This independence is indeed observed if we compare the curves $1'$ and 3 in Fig. 10a in the region $z < 1\, km$ where the PCBs are not sufficiently modified by the turbulent atmosphere. The high level of SI, $\sigma_s^2(z)$, calculated for the set $\{I_m^l(S = 1cm^2)\}$ is a consequence of the strong PM. (See Fig.10b). Since the states of the PM are statistically independent, the averaging over its $M$ states results in much lower SI, $\hat{\sigma}^2$.

After approaching the critical distance ($z > Z_{cr} = 1.65 km$ -see explanation for (7)) the independence of the signals, $I_m^l$, is suppressed by the atmosphere, and for $z > 5km$, the ratio $\sigma_s^2/\hat{\sigma}^2 \sim 1.3 \ll M = 16$.

Recall that in the laboratory experiment [12], PCBs with a small initial coherence radius were successfully used. But the effect of suppression of the SI was achieved in [12] in conditions for which beam wandering was absent after the atmospheric modulator (diffuser), which simulated a turbulent atmosphere. (The size of the microlenses of the AM was much smaller than the radius of PCB.) In a realistic atmosphere the situation is just the opposite. The spectrum, $\Phi_{n'}(q)$, (see Eq. (6)) includes harmonics with wavelengths, $\Lambda > R_s$ ($\Phi_{n'}(q = 2\pi/\Lambda)$ increases nonlinearly with $\Lambda$), and beam wandering prevents the reduction of SI. (See Fig.10*a*.)

### 3. Reduction of wandering by using optical vortices in a turbulent atmosphere

Here we demonstrate SI reduction by utilizing the partial suppression of the beam wandering for PCBs in the form of optical vortices (OVs). For convenience, we first present some known results for PCBs with OVs [13]. One way to create an OV is shown in Fig. 11 [19,20,21]. A Gaussian laser beam passes through a spiral phase mask which modulates the phase of the beam, $\varphi(x,y) = Arc\tan(y/x) \equiv \beta$, where $\beta$ is the angle of rotation around the direction of propagation, $z$. After passing through the mask, the optical field has the complex amplitude,

$$U(r, z = 0) = U_0 r \exp(-r^2/r_0^2) e^{i\beta}. \qquad (10)$$

(In general, $\varphi(x,y) = \pm j\beta$; where $j$ is an integer that is called the topological charge.)

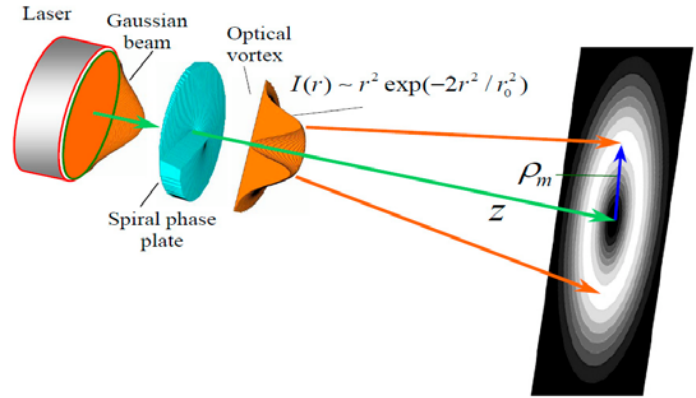

**Figure 11.** One possible realization of a PCB in the form of an OV [12,13]. The OV is shown here in longitudinal cross-section.

The solution of the paraxial equation (3) with the boundary condition (10) has the form [22]:

$$U(r, \beta, z) = \frac{U_0}{w(\tilde{z})^2} r \exp\left(-\frac{r^2}{r_0^2 \cdot w(\tilde{z})^2}\right) \exp[i\xi(r,\beta,\tilde{z})],$$

$$\tilde{z} = z/z_R, \ w(\tilde{z}) = \sqrt{1+\tilde{z}^2}, \quad \xi(\rho,\beta,\tau) = 2\arctan\tilde{z} - \tilde{z}\frac{r^2}{r_0^2 \cdot w(\tilde{z})^2} + \beta. \qquad (11)$$

The laser beam described by (11) is remarkable in the sense that its wave front is a helical surface with field intensity equal to zero at the axis of propagation, $z$. In any transverse plane, the field intensity is distributed according to the axially symmetric light circle shown in Fig. 11. The maximum of intensity occurs on the circumference with radius,

$$\rho_m = r_0 \frac{w(\tilde{z})}{\sqrt{2}}. \qquad (12)$$

The dark central area in Fig. 11 is caused by the initial helical phase perturbation, $\varphi(r) = \beta$. As a result of this perturbation, the "current streamlets" of the electromagnetic energy are the helical lines, which enclose the vortex axis.

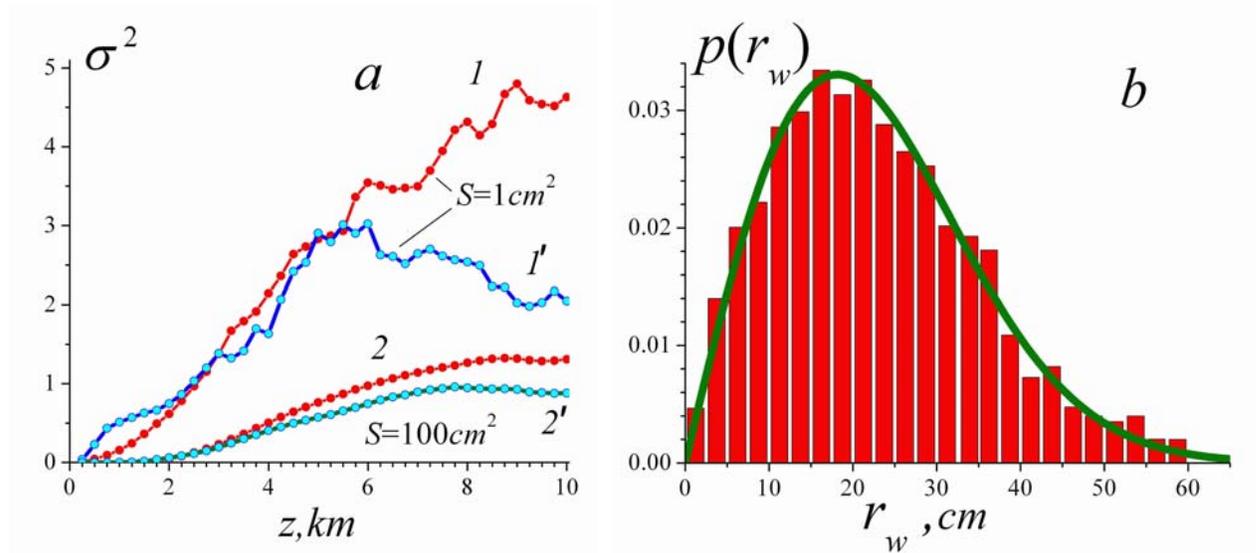

**Figure 12.** $a$ – Comparison of SIs for coherent Gaussian beams (curves $1, 2$) and an optical vortex (curves $1', 2'$). In both variants, $r_0 = 2 cm$ and averaging was performed over 5000 atmospheric states, $L_{atm} = 5000$. $b$ - The distribution function, $p(r_w; z=D)$, for the optical vortex; $p(r_w) \sim r_w \exp\left[-r_w^2/(2r_{w,\max}^2)\right]$, $r_{w,\max}^2 \approx 330 cm^2$.

Numerical simulations of the propagation of the OVs in turbulent atmospheres were done in [23]. It was discussed in [23-28] that the OV is less influenced by turbulence: for an OV the value of the SI is smaller, approximately by a factor of 1.5 [23], compared with the case of a coherent Gaussian beam. The physical processes responsible for this behavior were not discussed in [23-28].

Below we show that an OV can compensate its own wandering (which leads in the case of a Gaussian beam to an increased SI, as indicated above). Because of this so-called auto-compensation, we may obtain for the OV a lower SI compared with that for a Gaussian beam. However, in order to decrease the SI, the parameter, $r_0$, should be chosen in a way that depends on both the index of turbulence $C_n^2$ and the distance to the detector, $z$. Analysis of this phenomenon is essential for reducing the SI. (See sections 4 and 5 below.)

The mechanism of auto-compensation is rather simple. First, consider the propagation of an OV through a homogeneous atmosphere. Suppose that $r_0 = 2cm$. In this case $z_R = 810.7m$ and for $z = 10km$ we have $\rho_m \approx 17.5cm$. Qualitatively one can conclude that in turbulent atmospheres the deviation of the beam in any direction from the z-axis by $r_w \sim 15-20cm$ should not cause any considerable variation of the signal, $I^1$. For these deviations, part of the intensity from the luminous circle (related to the unperturbed OV) falls on the detector anyway. The data in Fig. 12 confirm quantitatively our assumptions. As one can see, the SI for the OV decreases only in the region $z > 6km$. This observation has the following explanation. One can calculate the function: $\Delta(z) = \rho_m(z) - r_{w,\max}(z)$ (where $r_{w,\max}$ is the point at which the distribution function, $p(r_w)$ reaches its maximum; $r_{w,\max}(z=10km) \approx 18cm$, see Fig. 12b.) It is clear that for sufficiently large values of $\Delta(z)$, auto-compensation does not appear. The maximal compensation effect could be achieved if $\Delta(z) \approx 0$. In our case, $\Delta(z)$ increases for small $z$ and approaches its maximum (approximately $4cm$) at $z \approx 4km$. Then it decreases, approaching its smallest value $\Delta(z=10km) \approx 0.5-1cm$. The SI $\sigma^2(z; S = 1cm^2)$ has a similar spatial behavior for the OV.

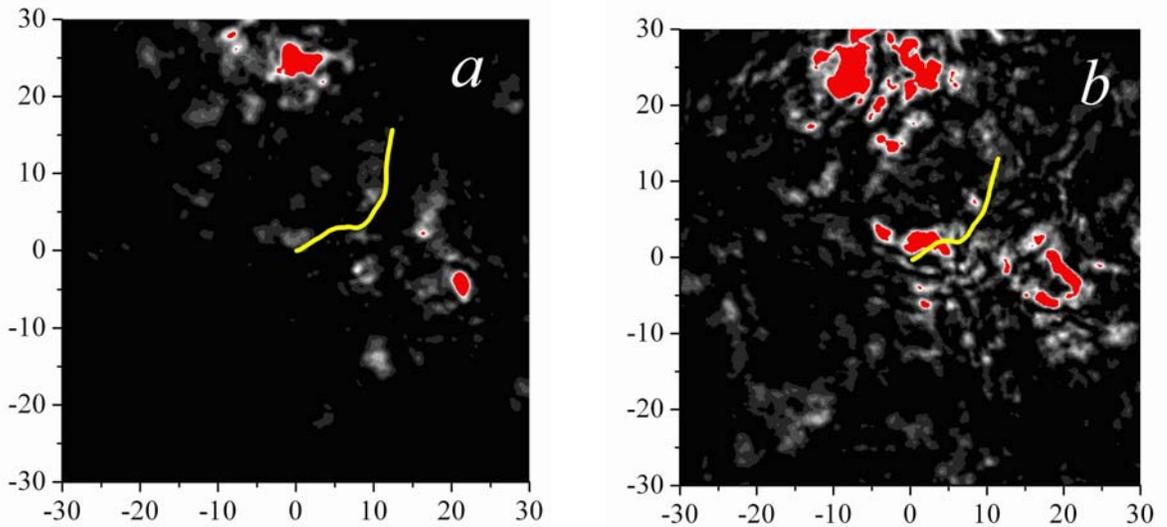

**Figure 13.** Distribution of the dimensionless intensity, $\tilde{I}(x, y, z = 10km)$, in the region $-30cm < x < 30cm$, $-30cm < y < 30cm$; $0 \leq \tilde{I}(x,y) \equiv I(x,y)/I_{\max}(x,y) \leq 1$. a- for a Gaussian beam, b- for an OV. Red indicates those regions with $\tilde{I}(x,y) \geq 0.2$ and yellow indicates the trajectories of $\vec{r}_w(z)$. Both distributions were calculated for the same atmospheric state.

So, the OV can only compensate those beam deviations, $r_w(z)$, that are close to the value of $\rho_m(z)$. This effect of the *partial* auto compensation is most effective when $\rho_m(z) \approx r_{w,\max}(z)$.

The effect of auto-compensation is demonstrated in Fig. 13. In contrast to the Gaussian beam (Fig. 13*a*), part of the OV (after a deviation by the atmosphere of the initial ring-shaped distribution of intensity) hits the detector located in the region with the center at $r = 0$. (See Fig. 13*b*.)

One might conclude that to achieve of auto-compensation, it is sufficient to create only an initial ring-shaped distribution of intensity (for example, by passing a beam through a mask with an inhomogeneous transparency). However, this action will not provide the desired result. Suppose, that in the distribution (10) we neglected the phase modulation factor, $e^{i\beta}$. For this hypothetical beam (HB), we have the distribution of intensity, $\tilde{I}(x,y,z=10km)$ (see Fig. 14) for the atmospheric state used in Fig. 13. This distribution is difficult to distinguish visually from that shown in Fig. 13a, calculated for Gaussian beam. This similarity is related to the fact that due to diffraction, the HB rather quickly transforms into a Gaussian-like beam (at $z \approx 500m < Z_{cr} \approx 1600m$). At the same time, the orbital momentum of the OV prevents this transformation.

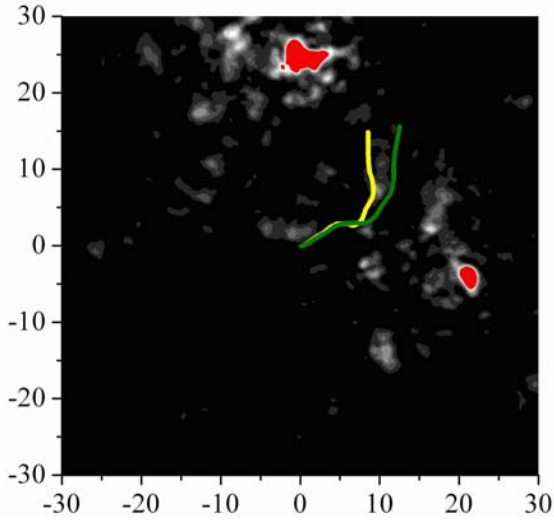

**Figure 14.** Distribution of dimensionless intensity, $\tilde{I}(x,y,z=10km)$ for the HB with initial distribution, $U(x,y) = U_0 r \exp(-r^2/r_0^2)$. Here, the phase modulation factor, $e^{i\beta}$, is excluded and $r_0 = 2cm$. Yellow indicates the trajectory, $\vec{r}_w(z)$; green indicates (for comparison) the trajectory of the center of the Gaussian beam. (See Fig. 13.*a*) Red indicates $\tilde{I}(x,y) > 0.2$.

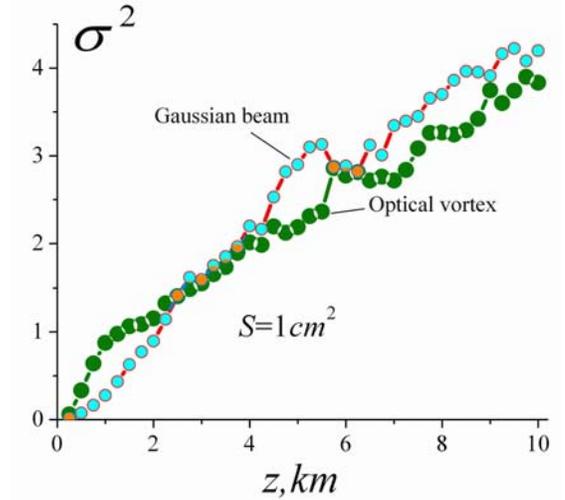

**Figure 15.** $\sigma^2(z)$ for a Gaussian beam and for an OV. In both variants, $r_0 = 5cm$. Partial auto-compensation of wandering for this OV almost does not appear. In the region $z < 2km$, $\sigma^2_{OV} > \sigma^2_{GB}$ due to (i) the low intensity of the disturbed OV on the axis results in signals, $I^l$, that are of order of their variations, $\delta I^l(z) = I^l(z) - \langle I^l(z) \rangle_l$ and (ii) significant fragmentation of the OV, which is inherent for the wider beam.

We conclude that to produce auto-compensation, the parameter, $r_0$, must correspond to the index of turbulence, $C_n^2$. Will the auto-compensation be effective in this case for an OV with $r_0 = 5cm$ ($C_n^2 = 2.5 \times 10^{-14} m^{-2/3}$, $z_R = 5067m$)? For $z = 10km$, we have $\rho_m = 11cm$. Due to the large difference between $\rho_m(z=10km)$ and $r_{w,\max}(z=10km) \approx 18cm$, the effect of auto-compensation is not achieved. (See Fig. 15.)

### 4. Suppression of scintillations by averaging over the angles of the laser beam

The reduction of the SI achieved by auto-compensation is not sufficient for achieving gigabit data rates and long-distance optical communications. (See Fig.12*a*.) We used a PCB, but for each atmospheric state, $l$, this beam (in fact, an OV) had a single state, without the consequent averaging of intensities, $I_m^l$, over the subscript, $m$ ($M=1$). Following a mentioned above idea of suppressing SI by the method of compensation of wandering, we present below the next step of the procedure.

For a non-turbulent atmosphere, we replace the homogeneous ring-shaped distribution of intensity in the detector plane by a set of separated in time coherent beams. These beams are oriented at an angle, $\theta$, to the direction of propagation, $z$, and they are homogeneously distributed in the azimuthal angle, $\beta$. The angle, $\beta$, takes discrete values, $\beta_j = \Delta\beta \cdot j$, $\Delta\beta = \dfrac{2\pi}{J}$, where $J$ is an integer. For each angle, $\beta_j$, one can observe a light spot at the screen positioned perpendicular to the z-axis. As the angle $\beta$ changes, the center of this spot moves discretely on the circumference of a circle with a radius of $\rho \approx z\theta$. We assume that the interval of this time series, $\tau$, is much smaller than the characteristic fluctuation time of the atmospheric state: $\tau_{atm} \sim 10^{-3}$ sec : $\tau \ll \tau_{atm}$. We will also call these laser beams PCBs because the variation of direction of coherent beam corresponds to changing its phase at the plane $z=0$: $\varphi_j(\vec{r}, z=0) = \vec{a}_j \cdot \vec{r}$, $|\vec{a}_j| = k\theta$, $k$ is the wave vector[*].

Generally, the angle $\theta$ must assume several discrete values, $\theta_i$, $i=1,2,...,I$, for the SI to be reduced sufficiently. Then, for each state of the turbulent atmosphere this set of PCBs is produced by means of phase modulation:

$$\varphi_{ij}(\vec{r}, z=0) = \vec{a}_{ij} \cdot \vec{r}, \quad i=1,2,...,I; \ j=1,2,...,J. \tag{13}$$

In (13) the modulus of $\vec{a}_{ij}$ is a function that depends only on the index $i$, and the angle $\theta_i \approx a_{ij}/k$.

Thus, averaging of intensity over all $M = I \cdot J$ states of the PCB,

$$\hat{I}^l = \frac{1}{M}\sum_{i,j} I_{ij}^l, \tag{14}$$

---

[*] The phase mask $\varphi(x,y,0) = \vec{a}\cdot\vec{r}$ transforms the flat wave front of a Gaussian beam, $z=0$, to the plane $zk + a_x x + a_y y = 0$ and the beam propagates along the normal to the plane, $\vec{n}$, which has components $\dfrac{a_x}{\sqrt{a^2+k^2}}, \dfrac{a_y}{\sqrt{a^2+k^2}}, \dfrac{k}{\sqrt{a^2+k^2}}$. So, $tg\theta = |a|/k \approx \theta$.

corresponds to averaging over the time interval, $\tau$.

Note that in this protocol, for each atmospheric state we use a set of PCBs (or a set $\{\vec{a}_{ij}\}$) that is chosen in advance, instead of using randomly generated sets.

Now we present the results of our numerical simulations for a beam with radius, $r_0 = 2cm$. As discussed above, the main contributions to the SI, $\sigma^2(z)$, of a coherent beam, are provided by the atmospheric states, $l$, for which $15cm < r_w^l(z = D = 10km) < 35cm$. To compensate the beam wandering in this interval, we define the set $\{\vec{a}_{ij}\}$ so that the angle, $\theta$, takes on three values: $\theta_i = \rho_i/D$, ("scattering parameter" $\rho_1 = 15cm$, $\rho_2 = 25cm$, $\rho_3 = 35cm$).

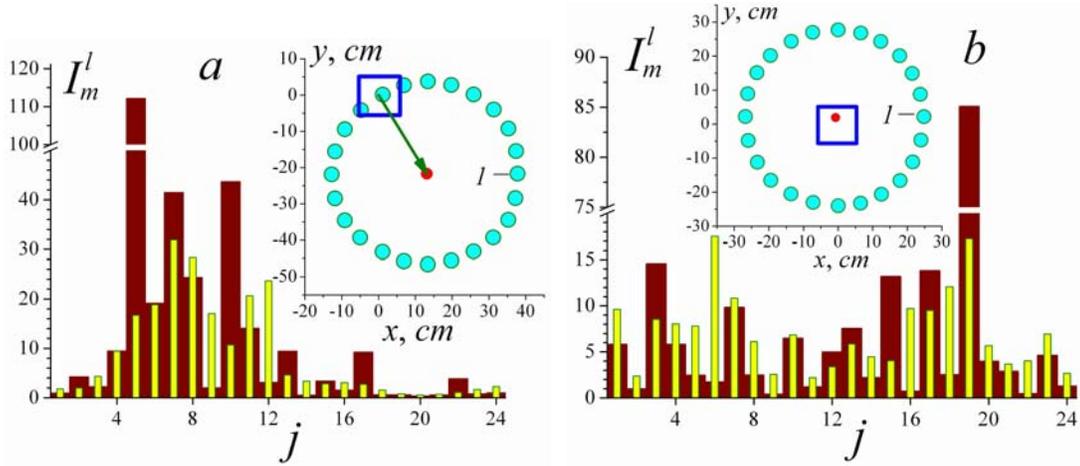

**Figure 16.** Results of propagation of a PCB with the scattering parameter, $\rho_2 = 25cm$, through the two atmospheric states, $l_1$ (Fig. 16a) and $l_2$ (Fig.16b); $z = 10km$. In these diagrams, the levels of intensity, $I_j^l$, are shown that are produced at the detector by each beam, separately (in conditional units: for a unit area of detector). Maroon indicates $S = 1cm^2$, yellow indicates $S = 100cm^2$. In the inserts: the blue square is the detector area, $100cm^2$; circles – centers of PCBs $\vec{r}_{w,j}^l(z = 10Km)$; numbering is anticlockwise starting from the most right point (point 1 in insert). Averaging was done over all 24 beam intensities: Fig. 16a- $\hat{I}_{s=1cm^2} = 12.87$ and $\hat{I}_{s=100cm^2} = 8.77$; Fig. 16b - $\hat{I}_{s=1cm^2} = 8.12$ and $\hat{I}_{s=100cm^2} = 7.15$.

The physical reasons that lead to suppression of the SI are demonstrated in Fig. 16. Namely, Fig. 16 shows a characteristic set of 24 beams (with scattering parameter $\rho_2 = 25cm$), which passed two random atmospheric states, $l_1$ and $l_2$. As Fig. 16 demonstrates, the centers of the PCBs, $\vec{r}_{w,j}^l$, $j = 1,2,...,24$; $l = l_1, l_2$, approximate a circumference with radius of the order of the scattering parameter, $\rho_2$. The center of this circumference (red points in inserts), is close the center of a coherent beam, $\vec{r}^{\,l}_{w,0}$, which has traversed through the same atmospheric states. The reason of this phenomenon is that the deviation of the beams from their initial $z$-directions is mainly due to large-scale inhomogeneities of the index of refraction, $n'$, with characteristic scales, $\Lambda$, that are much larger than the effective beam radius, $R_s (z \leq 10 km)$, and scattering parameter, $\rho_2$, in the case under consideration. So, the trajectories $\vec{r}^{\,l}_{w,j}(z)$ and $\vec{r}^{\,l}_{w,0}(z)$ are strongly correlated.

- In the first case ($l = l_1$, Fig. 16a), one can clearly see a correlation of intensity, $I^{l_1}_j$, with remoteness, $r^{l_1}_{w,j}$, the center of the $j$-th beam from detector. For $S = 100 cm^2$ the average relative deviation is: $\delta^2 = \left[ \left\langle \left(I^{l_1}_j\right)^2 \right\rangle_j - \left\langle I^{l_1}_j \right\rangle^2_j \right] / \left\langle I^{l_1}_j \right\rangle^2_j \approx 1.2$.

- In the second case, ($l = l_2$, Fig. 16b), the variation of values $I^{l_2}_j$ is significantly smaller than for $l = l_1$ because $r^{l_2}_{w,j} \approx const$, the corresponding $\delta^2 \approx 0.35$. (Naturally, if we choose all states of the atmosphere that cause small deviations of coherent beam, $r^{l}_{w,0}$, each PCB will give on average an approximately equal contribution to the intensity, $\hat{I}^l$.)

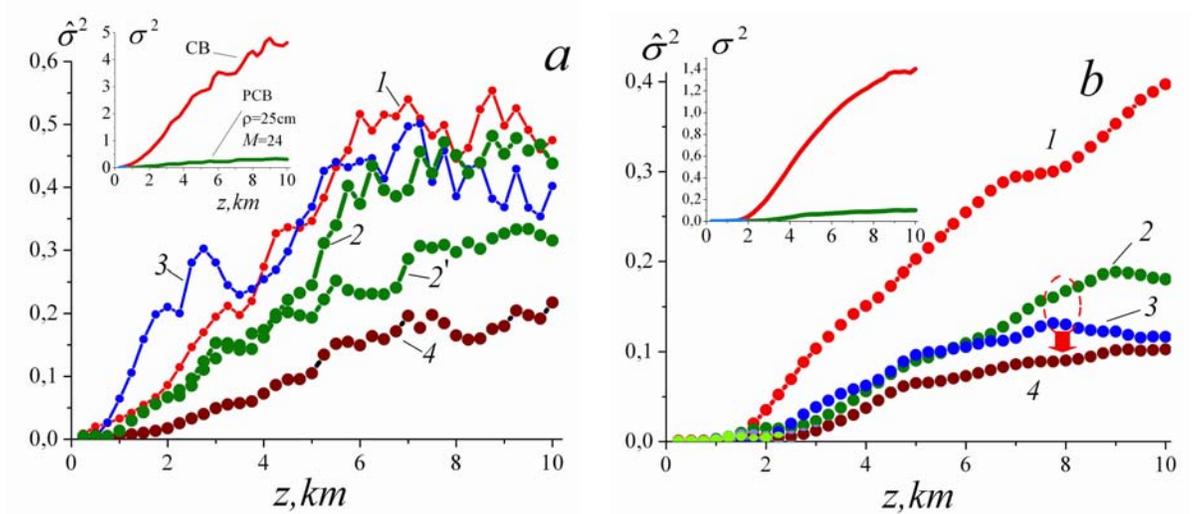

**Figure 17**. Dependence of the scintillation index on the distance, $z$, for detectors with two different areas. $a - S = 1 cm^2$. curves 1,2,3 - $\hat{\sigma}^2_{(1)}$, $\hat{\sigma}^2_{(2)}$, $\hat{\sigma}^2_{(3)}$ ($M = 12$), correspondingly; curve 4 - $\hat{\sigma}^2$ ($M = 36$); curve $2'$ - $\hat{\sigma}^2_{(2)}$ for parameters $J = 24$. In the insert: $\sigma^2(z)$ for a coherent beam (red) and for comparison $\hat{\sigma}^2_{(2)}(z; J = 24)$ (green). $b - S = 100 cm^2$. Curves 1,2,3 - $\hat{\sigma}^2_{(1)}$, $\hat{\sigma}^2_{(2)}$, $\hat{\sigma}^2_{(3)}$; curve 4 - $\hat{\sigma}^2(z)$, calculated for a set of PCBs beams for which a scattering parameter takes two values ($I = 2$), $\rho = 25$ and $35 cm$ ($M = 2 \times 12 = 24$ - data for curves 2 and 3 are combined). In the insert for comparison, $\sigma^2(z)$ for a coherent beam and $\hat{\sigma}^2_{(2)}$.

For $l = l_1$, those beams that pass close to the detector compensate the small contribution of those beams with large deviations from the detector. For $l = l_2$, the contribution of each of the 24 beams is moderate, but relatively uniform. As a result, the integral characteristics of intensity, $\hat{I}^l = \frac{1}{J}\sum_j I_j^l$, in both of these cases (see the caption in Fig. 16) and for other atmospheric states are practically equalized. That is the main mechanism responsible for the suppression of the SI: $\hat{\sigma}^2 \ll \sigma^2$.

In most simulations we have chosen $J = 12$. As shown below, $J = 12$ is optimal for suppression of SI for the chosen parameters ($C_n^2 = 2.5 \times 10^{-14} m^{-2/3}$, $r_0 = 2cm$, $D = 10km$, $S = 1-100 cm^2$). We calculated the statistical characteristics for the sets with one, two and three scattering parameters (in the last case $M = 3 \times 12 = 36$). The SI for the set with a single scattering parameter (one from $\rho_i = 15, 25, 35 cm$), we will call "partial" and define them by $\hat{\sigma}_{(1)}^2$, $\hat{\sigma}_{(2)}^2$, $\hat{\sigma}_{(3)}^2$, depending on chosen value of $\rho_i$.

We now discuss the results for detectors with $S = 1 cm^2$. (See Fig. 17a.) As one can see, even the partial SIs, $\hat{\sigma}_{(i)}^2(z = D)$ (curves 1, 2, 2', and 3) are 10 to 14 times smaller than the SI for a coherent beam. Averaging over all 36 possible states of the PCB is even more effective: $\sigma^2/\hat{\sigma}^2 \approx 20$. It is to be noticed that the set of separated in time single beams whose centers are on the circle with radius, $\rho(D)$, is much more effective in reducing the SI than the OV with the same radius, $\rho_m(D) = \rho(D)$. The reason for this is that in the first case, the interference of the fields, $U_j^l$, created by the single beams at the detector is excluded, $\hat{I}^l = \frac{1}{J}\sum_{j=1}^{J}|U_j^l|^2$. In the case of OV, which can be mentally divided into $J$ equal parts at the plane $z = 0$, the interference is not avoided at the detector, $\hat{I}^l = \frac{1}{J}\sum_{j=1}^{J}|U_j^l|^2 + \frac{1}{J}\sum_{j,k=1}^{J}U_j^l U_k^{*l}$. The second term describes the "extra noise" which diminishes the effectiveness of the OV.

For the scattering parameter, $\rho_2 = 25 cm$, the SI, $\hat{\sigma}_{(2)}^2$, was also calculated for $J = 24$ (curve 2' in Fig. 17a). Increasing $J$ by a factor two appears only for the region $z > 5 Km$. Indeed, for an arbitrary scattering parameter, $\rho_i$, a threshold should exist, $J_{sat}(z)$, above which saturation occurs in the dependence, $\hat{\sigma}_{(i)}^2(J)$: $\hat{\sigma}_{(i)}^2(z; J > J_{sat}) \approx const$. This is due to the fact that $I_{ij}^l$, $j = 1, 2, ..., J$, for a given $z$, become less statistically independent in the averaging (14) due to decreased gaps between trajectories of the beams, $\vec{r}_{w,2j}^l(z)$. The corresponding correlations first reveal themselves for smaller distances where the density of the beam trajectories ($\vec{r}_{w,ij}^l(z)$, $j = 1, 2, ..., J$) is higher. It is physically clear that the value, $J_{sat}$, depends not only on $z$, but also on both (i) the area of the detector, $S$, and (ii) the scattering parameter, $\rho$:

(i) As $S$ increases, the correlations between the average intensities at the detector, $I_{ij}^l = \frac{1}{S}\int_S I_{ij}^l(x,y)ds$, with different indices, $j$, increases also and $J_{sat}$ decreases.

(ii) As $\rho$ increases, the correlations between the intensities, $I_{ij}^l$, decreases, because the beams pass through atmospheric channels that are physically distinct, with different optical properties and $J_{sat}$ increases.

*Conclusion*. By implementation of the supposed method of reducing the SI, the total number of beams, $M$, is limited by the frequency, $f = 1/\tau_{ph}$, of the phase modulator: $M \leq \tau/\tau_{ph}$ ($\tau$ is the length of the time series required to achieve high-rate optical communications and $\tau_{ph}$ is the time of creating the single phase mask, $\varphi_{ij}(x,y)$). So, to achieve the maximal suppression of the SI for given $M$, the set of scattering parameters, $\{\rho_i\}$, and the number of beams for each of these parameters, $J_i$, must be optimized (by the condition that $\sum_i J_i = M$).

We now discuss the results presented in Fig. 17b. As one can see, a significant suppression of the SI is achieved. Here we concentrate on understanding the statements (i) and (ii).

The scattering parameter, $\rho_1 = 15cm$, is not as effective as the scattering parameters $\rho_2$ and $\rho_3$ for $S = 100cm^2$. (For $S = 1cm^2$, all three scattering parameters were practically equivalent.) The reason is that for the scattering parameter $\rho_1$ the value $J_{sat} < J = 12$, and only a subset of the beams is effective in the averaging (14).

We now provide an estimate for $J_{sat}$ for arbitrary index $i$ (*i.e.*, considering only a single scattering parameter $\rho$), assuming for some number, $J$, the intensities, $I_j^l$, are calculated ($l = 1, 2, ..., L_{atm}$, $j = 1, 2, ..., J$). The value $\hat{\sigma}^2 = \frac{\langle(\hat{I}^l)^2\rangle - \langle\hat{I}^l\rangle^2}{\langle\hat{I}^l\rangle^2}$ for the averaged signal, $\hat{I}^l = \frac{1}{J}\sum_{j=1}^J I_j^l$, can be written in a better form. For each subset $\{I_j^l\}$, $l = 1, 2, ..., L_{atm}$, we introduce the average value, $I_j^{av} \equiv \langle I_j^l\rangle_l$, (accordingly, $I_j^l = I_j^{av} + \delta I_j^l$) and the corresponding partial scintillation index, $\sigma_j^2 = \frac{\langle(I_j^l)^2\rangle_l - \langle I_j^{av}\rangle^2}{\langle I_j^{av}\rangle^2}$, where averaging was done over the atmospheric states, $l$. Then

$$\hat{I}^l = \frac{1}{J}\sum_{j=1}^J (I_j^{av} + \delta I_j^l), \quad \langle\hat{I}^l\rangle = \frac{1}{J}\sum_{j=1}^J I_j^{av}, \quad \sigma_j^2 = \frac{\langle(\delta I_j^l)^2\rangle_l}{\langle I_j^{av}\rangle^2}, \tag{15}$$

and $$\left\langle (\hat{I}^l)^2 \right\rangle = \left\langle \hat{I}^l \right\rangle^2 + \frac{1}{J^2}\sum_{j=1}^{J}\left\langle (\delta I_j^l)^2 \right\rangle_l + \frac{1}{J^2}\sum_{j,k=1}^{J}\left\langle \delta I_j^l \delta I_k^l \right\rangle_l . \tag{16}$$

Taking into consideration (15) and (16), the new formula for $\hat{\sigma}^2$ is

$$\hat{\sigma}^2 = \hat{\sigma}_0^2 + \left(\sum_{j,k=1}^{J}\left\langle \delta I_j^l \delta I_k^l \right\rangle_l\right) \Big/ \left(\sum_{j=1}^{J} I_j^{av}\right)^2, \quad j \neq k, \tag{17}$$

where $\hat{\sigma}_0^2 = \sum_{j=1}^{J} I_j^{av} \sigma_j^2 \Big/ \left(\sum_{j=1}^{J} I_j^{av}\right)^2$ is the "optimal" SI if the set of PCBs creates statistically independent intensities at the detector, $\left\langle \delta I_j^l \delta I_k^l \right\rangle_l = 0$, $j \neq k$. (Here we consider only a single scattering parameter, $\rho$.)

The calculation of the both SI $\hat{\sigma}^2$ and $\hat{\sigma}_0^2$ can be easily done using the datagenerated in our calculations. The effectiveness, $\eta$, of the chosen number, $J$, can be estimated as $\eta = \hat{\sigma}_0^2/\hat{\sigma}^2 \leq 1$, accordingly

$$J_{sat} \approx J \hat{\sigma}_0^2 / \hat{\sigma}^2 . \tag{18}$$

Next we present an example that demonstrates correctness of (18). For $J = 24$, $\rho = \rho_2 = 25cm$ and $S = 100cm^2$ we obtained $\hat{\sigma}_{(2)}^2(J = 24; z = 10km) \approx 0.18$, for the corresponding optimal SI, $\hat{\sigma}_0^2 \approx 0.08$. According (18), $J_{sat} \approx 11$. Actually, the calculated dependence $\hat{\sigma}_{(2)}^2(z; J = 12)$ (see Fig.17b, curve 2) essentially coincides with $\hat{\sigma}_{(2)}^2(z; J = 24)$. (Incidentally, in the last case $\hat{\sigma}_0^2(J = 12) \approx 0.16$ and this again results in $J_{sat} \approx 12 \cdot 0.16 / 0.18 \approx 11$).

For $\rho_1 = 15cm$ ($J = 12$, see Fig.17b, curve 1), the ratio $\eta = \left(\hat{\sigma}_0^2 / \hat{\sigma}_{(1)}^2\right)_{z=10km} \approx 12/39$. Due to its reduced effectiveness, curve 1 in Fig. 17b is reproduced almost exactly for $J = 6$. Because of this, curve 4 in Fig. 17b was calculated by averaging only over two scattering parameters ($\rho_2$ and $\rho_3$). The additional 12 beams with scattering parameter $\rho_1$ practically did not change the result. Finally, the present value of $\eta$ for $S = 1cm^2$ ($z = 10km$) is $\eta(\rho_1; J = 12) \approx 0.64$, $\eta(\rho_2; J = 12) = \eta(\rho_3; J = 12) \approx 1$; $\eta(\rho_2; J = 24) \approx 0.77$.

Finally, when realizing the described above method for suppression of scintillations, the set of angles, $\theta_i$, and values, $J_i$, must be optimized taking into account the strength of atmospheric turbulence, size of detector and distance of propagation. To effectively manage the phase modulator for a specific situation, one must first produce an appropriate database, $\varphi_{ij}(x,y)$, by numerical modeling the propagation of laser beams through turbulent atmospheres.

## 5. Suppression of scintillations by using an asymmetric optical vortex

The generation of a set of PCBs that effectively reduces the SI (See Section 4.) faces significant technical problems for high-data-rate optical communication, because the transmission of information at rates above $1 Gbit/s$ requires a phase modulator, $\varphi_{ij}(x,y)$, with a frequency above $10^{10} Hz$. Below we discuss one method to solve this technical problem.

Consider the properties of a specific PCB that is a superposition of an optical vortex (OV) and a Gaussian beam. The axes of these beams coincide with the $z$-axis. The amplitudes of the OV and the Gaussian beam at the plane $z = 0$ are

$$U_{ov}(r,0) = A \times \frac{r}{r_{v0}} \exp\left(-\frac{r^2}{r_{v0}^2}\right) e^{i\beta + i\omega_v t} \text{ and } U_g(r,0) = B \times \exp\left(-\frac{r^2}{r_{g0}^2}\right) e^{i\omega_g t}. \tag{19}$$

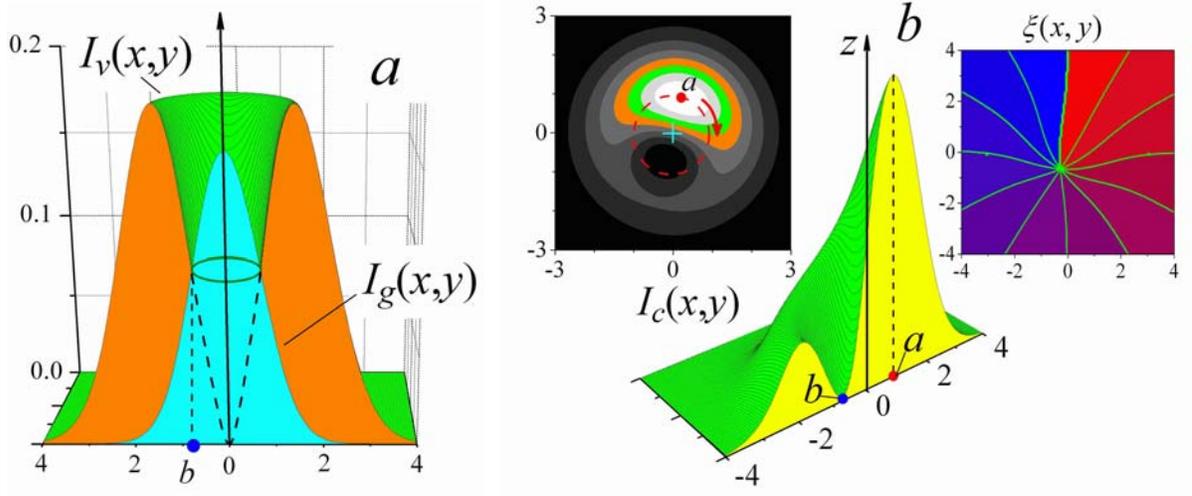

**Figure 18a** – Distribution of intensity of the OV, $I_v(x,y)$, and of the Gaussian beam, $I_g(x,y)$, in the plane, $z = 0$. (Distances are shown in cm.) $A = 1$, $B = 0.4$, $r_{0v} = 2 cm$, and $r_{0g} = r_{0v}/\sqrt{2}$; for point $b$, $r_b \approx 0.71 cm$.

**Figure 18b** – The result of the superposition of the two beams: a cross-section through the line $ab$ of the distribution of intensity of the CB, $I_c(x,y)$; for the point of maximal intensity, $a$, $r_a \approx 0.9 cm$. In the left insert, contours of the rotating distribution, $I_c(x,y,t)$, at an instant of time. In the colored areas, the intensity $I_c(x,y,t) > 0.5 \cdot I_{c,\max}(x,y,t)$. The right insert is a diagram of the phase of the CB, $\xi = Arc\tan\left[\text{Im}(U_c)/\text{Re}(U_c)\right]$.

The parameters of the beams (the amplitudes, $A$ and $B$, and the radii, $r_{v0}$ and $r_{g0}$) are chosen so that the distribution of the intensity of the Gaussian beam, $I_g(x,y,z=0)$, is imbedded in the "funnel" provided by the distribution of intensity of the OV, $I_v(x,y,z=0)$. (See Fig.

18a). In this case, there is a circle with radius, $r_b$, on which $|U_{ov}(r_b)| = |U_g(r_b)|$. As the phase of the OV varies from $0$ to $2\pi$ on this circle, the point, $b$, exists at which the phases of two beams are opposite and $U_c(r_b, \beta, z=0) \equiv U_{ov}(r_b, \beta, z=0) + U_g(r_b, \beta, z=0) = 0$. At the radially opposing point, $a$, the intensity of the combined beam (CB), $I_c(r_a, \beta+\pi, z=0) = |U_c(r_a, \beta+\pi, z=0)|^2$, is maximal. The result of the superposition of the two beams at time $t=0$ is shown in Fig. 18b. The distribution of the phase $\xi(x, y, z=0)$ demonstrates that the CB is an optical vortex with a non-symmetrical distribution of intensity. (See Fig. 18b, right insert.)

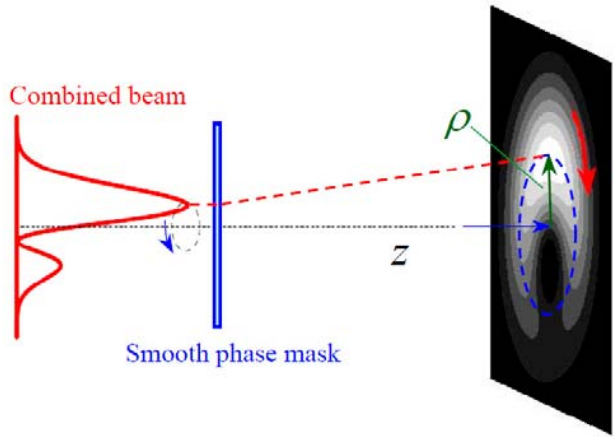

**Figure 19.** A scheme for achieving a required scattering parameter of CB in the region of detector.

We consider here the case in which the frequencies, $\omega_{ov}$ and $\omega_g$, differ by the amount, $\delta\omega = |\omega_{ov} - \omega_g| \ll \omega_{ov}, \omega_g$. In this case, the asymmetric distribution, $I_c(x, y, z=0)$, rotates around the $z$-axis with the frequency, $\delta\omega$, anticlockwise (if $\omega_{ov} > \omega_g$): the maximum of the intensity occurs at $r = r_a$, with the azimuthal angle, $\beta_a = const - \delta\omega \cdot t$. The value, $\delta\omega$, could be chosen in a desired region (e.g. $\delta\omega \sim 10^{10} Hz$ and above).

In a *homogeneous atmosphere*, the maximum of the intensity of the CB and its center, $\left( \vec{r}_c(z) = \dfrac{\int I_c(x,y,z)\vec{r}ds}{\int I_c(x,y,z)ds} \right)$, deviates from the $z$-axis due to diffraction spreading. In the plane of the detector perpendicular to the $z$-axis, both points move on concentric circumferences with radii of $r = r_a(z)$ and $r_c(z)$ ($r_a > r_c$). Thus, a CB is a continuous series of PCBs, oriented in the direction of detector with scattering parameter, $\rho(z) \equiv r_a(z)$. (From two characteristic values, we have chosen the more physically visible.)

In a turbulent atmosphere, this set of PCBs could compensate the wandering. (See Section 4.) This leads to a reduced SI for the averaged intensity,

$$\hat{I}^l = \frac{1}{2\pi} \int_0^{2\pi} I^l_{\beta_a} d\beta_a .\qquad(20)$$

In (20) the averaging was done over one period of rotation of the CB, $0 \le \beta_a \le 2\pi$, $\beta_a$ is the azimuth of the maximum of $I_c(x, y, z=0)$.

For useful suppression of the SI, we used an additional and important element in our numerical model–the phase mask, $\varphi_0(x, y, z \approx 0)$, (independent of time. See Fig. 19.) The need of this mask is motivated by the following. For the given parameters of the CB (see Fig. 18) at $z = 10 km$, the scattering parameter is $\rho \equiv r_a(z = D) \approx 16 cm$ ($r_c(z = D) \approx 7.6 cm$, and the center of the OV is at $r_b(z = D) \approx 5 cm$). For these conditions, an effective compensation of the wandering of the beams with $r_w(z = 10 km) \approx 15 - 35 cm$ ($C_n^2 = 2.5 \times 10^{-14} m^{-2/3}$) cannot be achieved. Consequently, one must position a phase mask in the path of propagation of the CB. This mask can provide beam spreading larger than the diffraction widening. After passing through this mask, the scattering parameter, $\rho$, can be increased to the required value at the areas of the location of the detector. The optimal choice of the function $\varphi_0(x, y, z \approx 0)$ is a complicated problem. Averaging (20) over a continuous set of $\beta_a$ is equivalent to averaging over a discrete set $\beta_{a,j}$ ($j = 1, 2, ..., J$) for the proper choice of $J$. Maximal reduction of the SI can be achieved if the values $I^l_{\beta_{a,j}} \equiv I^l_j$ are statistically independent. Consequently the phase modulation, $\varphi_0(x, y, z \approx 0)$, should not increase the region of overlap of the distribution of intensity of the CB, $I_{c,m}(x, y)$ and $I_{c,n}(x, y)$, for different moments of time (for different angles $\beta_{a,j}$, $j = m, n$). This condition is not satisfied, for example, for the radial modulation, $\varphi_0(r) = -\gamma r^2$.

In our case, we have chosen a phase modulation in the form of a cone of revolution

$$\varphi_0(r) = 4\frac{r}{r_{v0}}, \tag{21}$$

in which $r_{v0} = 2 cm$ is the radius of the OV.

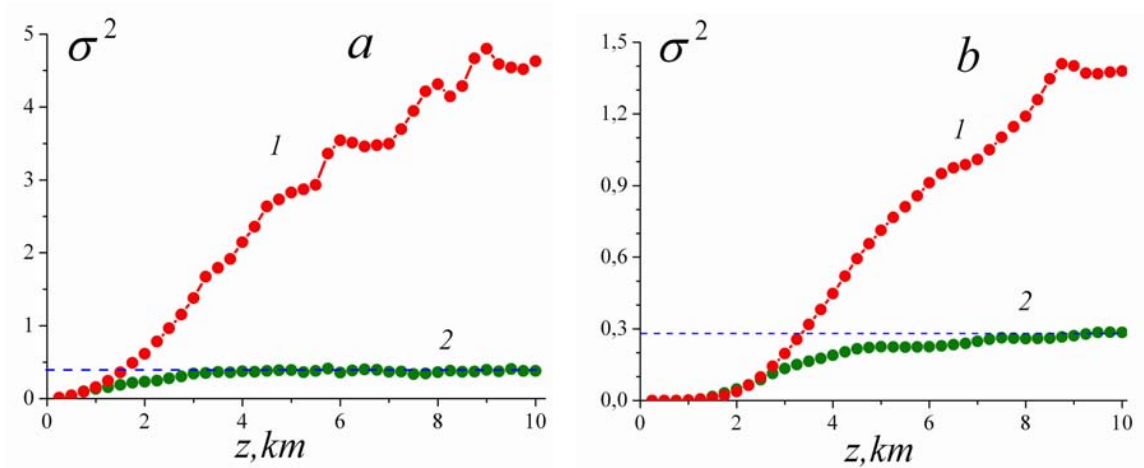

**Figure 20.** Comparison of SIs calculated for a coherent Gaussian beam (curve 1, $r_0 = 2 cm$) and for a CB (curves 2, $r_{v0} = 2 cm$, $r_{g0} = r_{v0}/\sqrt{2} \approx 1.4 cm$). $a$ -area of detector $S = 1 cm^2$; $b$ - $S = 100 cm^2$.

This modulation provides a scattering parameter, $\rho(z=10km) \approx 35cm$. The results of our numerical simulations are presented in Fig. 20 (with $J=12$).

Note that a significant decrease of the SI was achieved without a time-dependent phase modulation (13). (Compare these results with results presented in Fig. 17.) However, the possibilities of our approach are not limited by the above example. The values of the parameter of effectiveness, $\eta$, for the chosen number, $J$, are: $\eta \approx 0.45$ ($S=1cm^2$) and $\eta \approx 0.3$ ($S=100cm^2$). In this last variant, it was sufficient to choose $J=4$ in order to reproduce the result presented in Fig. 20b. (i.e., $J_{sat} \approx 4$. See. Fig.21a.)

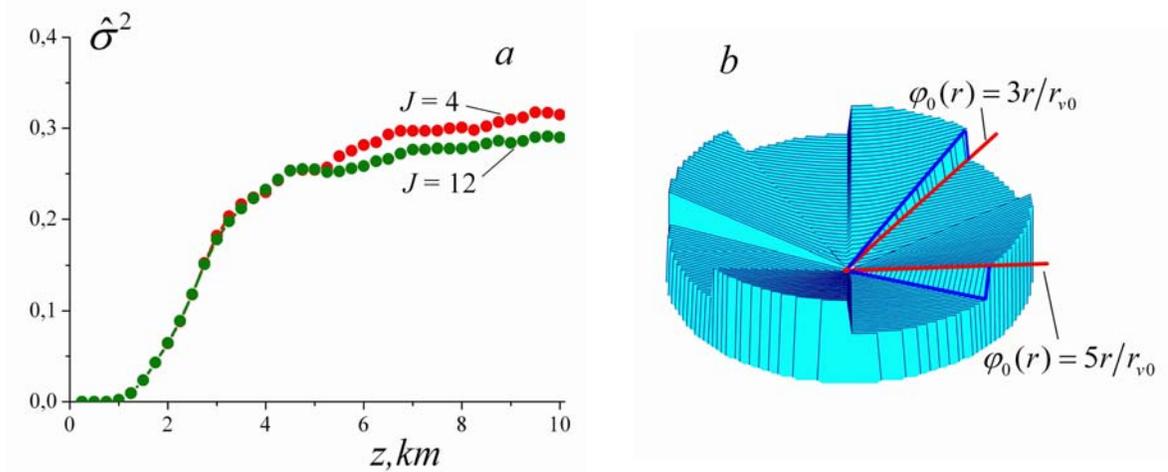

**Figure 21**. *a*-The z-dependence of $\hat{\sigma}^2(z; S=100cm^2)$ calculated for two different values of $J$. *b*- An example of an azimuthally asymmetrical phase mask that can result in an effective reduction of the SI.

This low value for $J_{sat}$ suggests that it might be possible to achieve an even more effective reduction of SI. For the intensities, $I^l_{\beta_a, j}$, to become statistically independent, we should use an azimuthally asymmetrical phase mask, $\varphi_0(x,y)$. For example, using the phase mask shown in Fig. 21b results in a reduction of the SI by a factor of ~ 2.4 ($J=16$) relative to the results in Fig.20 and Fig.21a.

### 6. Suppression of scintillations for strong turbulence

Below we present the results of numerical simulations for the case of relatively strong turbulence, $C_n^2 = 10^{-13} m^{-2/3}$. (This turbulence is four times larger than the value that was used previously, $C_n^2 = 2.5 \times 10^{-14} m^{-2/3}$.) All calculated values and parameters associated with the turbulence strength, $C_n^2 = 2.5 \times 10^{-14} m^{-2/3}$, are indicated by the subscript, $A$; and for $C_n^2 = 10^{-13} m^{-2/3}$ by the subscript, $B$. We now compare the statistical geometrical averages for coherent beam ($r_0 = 2cm$) for these two cases.

When the turbulence is increased by a factor of 4, the value, $\langle R^2(z > Z_{cr})\rangle$, also increases by a factor of 4 (in agreement with the analytical dependence (7); see Fig. 22). The wandering of the beam increases less significantly: $\sqrt{\langle r_w^2\rangle_B / \langle r_w^2\rangle_A} \approx 1.5$. (See the insert in Fig. 22.)

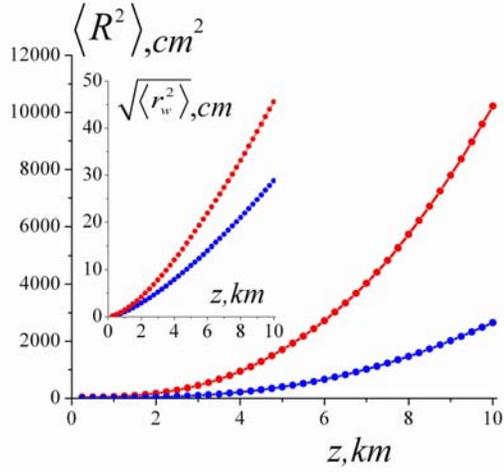 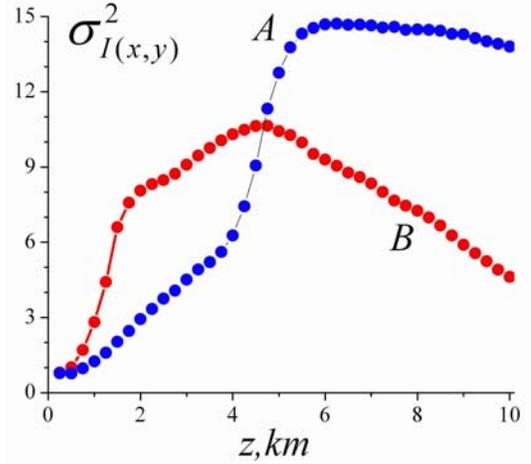

**Figure 22.** Radius squared and wandering of a coherent beam ($r_0 = 2cm$) as a function of the distance, $z$. Red - $C_n^2 = 10^{-13} m^{-2/3}$; blue - $C_n^2 = 2.5 \times 10^{-14} m^{-2/3}$.

**Figure 23.** Z-dependence $\sigma^2_{I(x,y)}(z)$ for (A) $C_n^2 = 2.5 \times 10^{-14} m^{-2/3}$ and (B) $C_n^2 = 10^{-13} m^{-2/3}$.

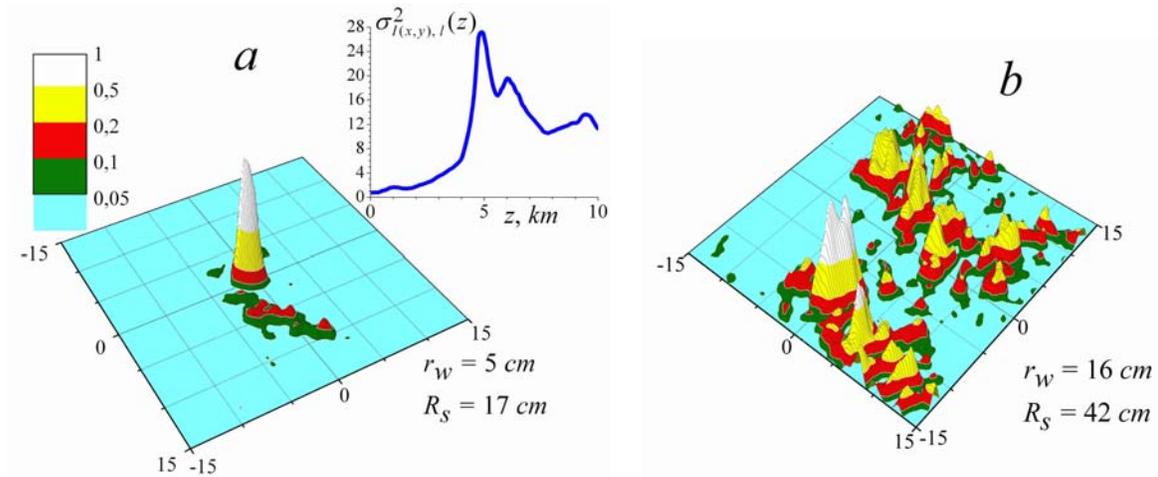

**Figure 24.** An example of the decay of a strongly fragmented beam into secondary fragments. Spatial distributions for dimensionless intensity of the beam, $\tilde{I}(x,y)$ ($0 \leq \tilde{I}(x,y) \leq 1$) are shown in the coordinate system connected to the center of the beam (which is shifted from the z-axis to the point, $\vec{r}_w(z)$). Distances in the xy-plane are shown in centimeters. $C_n^2 = 2.5 \times 10^{-14} m^{-2/3}$. a – $z = 5km$, b - $z = 10km$. In the insert, $\sigma^2_{\tilde{I}(x,y),I}(z)$ is shown.

To demonstrate the modifications in the spatial structures of the beam, we calculated $\sigma^2_{I(x,y)}(z)$ for the indicated values of $C_n^2$. (See Eqs. (8), (9) and Fig. 23.) Recall that the value, $\sigma^2_{I(x,y),l}(z)$, for a given atmospheric realization, $l$, characterizes the level of inhomogeneity of the intensity distribution in the beam cross section. For a given beam size, having fewer fragments in the distribution, $I^l(x,y)$, corresponds to increased $\sigma^2_{I(x,y)}$.

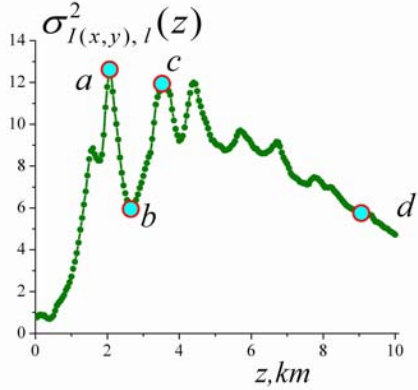

**Figure 25.** An example of "relaxation oscillations" in the level of beam fragmentation before decay into a set of small fragments for $C_n^2 = 10^{-13} m^{-2/3}$. The distributions, $\tilde{I}(x,y)$, for the points $a-d$ are presented in Fig. 26.

Note that the $z$-dependences in Fig. 23 are obtained as a result of averaging over all atmospheric states. For individual atmospheric states, fluctuations of the fragmentation patterns are observed even for weak turbulence. (See. Fig. 24.) For $C_n^2 = 10^{-13} m^{-2/3}$, the "relaxation oscillations" of the level of fragmentation often occur before the beam decays into a set of small fragments. (See. Fig. 25.)

The set of distributions of the beam intensity shown in Fig. 26 gives a visual presentation of the relation between the parameter, $\sigma^2_{I(x,y),l}(z)$, and the beam structure in the transverse cross section.

Finally, we present the distribution function, $p(r_w)$, and the statistical characteristics of the signal at the detector (see Fig. 27) for the case of strong turbulence.

By comparing the results presented in Figs. 22-27, we can understand the main characteristics of a coherent beam propagating through the most turbulent atmosphere:

**1**. For small distances ($z \leq 4km$ in Fig. 23), fragmentation of the beam significantly increases with increasing $C_n^2$. In this region, one can expect $\sigma_B^2(z) > \sigma_A^2(z)$.

**2**. For large distances ($z \geq 5km$ in Fig. 23), individual fragments of the beam decay into secondary fragments. The overlapping of the secondary fragments, due to their diffractive spreading and scattering in the turbulent media, leads to a decrease of the fragmentation parameter of the beam ($\sigma^2_{I(x,y)}(z)$) as the distance increases. This effect is more pronounced in the atmosphere with a strong turbulence (see curve $B$ in Fig. 23), and it results in decreasing SI with increasing distance.

**3**. For the larger turbulence strength:
- The ratio of the beam radius, $R_s(z)$, to $r_w(z)$ increases. (See Fig. 22.) For example, for $z = D$ $\langle R_s/r_w \rangle_A \approx 1.5$ but $\langle R_s/r_w \rangle_B \approx 2$ (averaged over all atmospheric states);

- The characteristic size of the speckle field (see Fig. 26d, $z = 10 km$) becomes comparable with the value, $l_0$ (where $l_0 = 6.5 mm$ is the inner scale of turbulence). This is significantly smaller than the size of the detector, $S = 100 cm^2$.

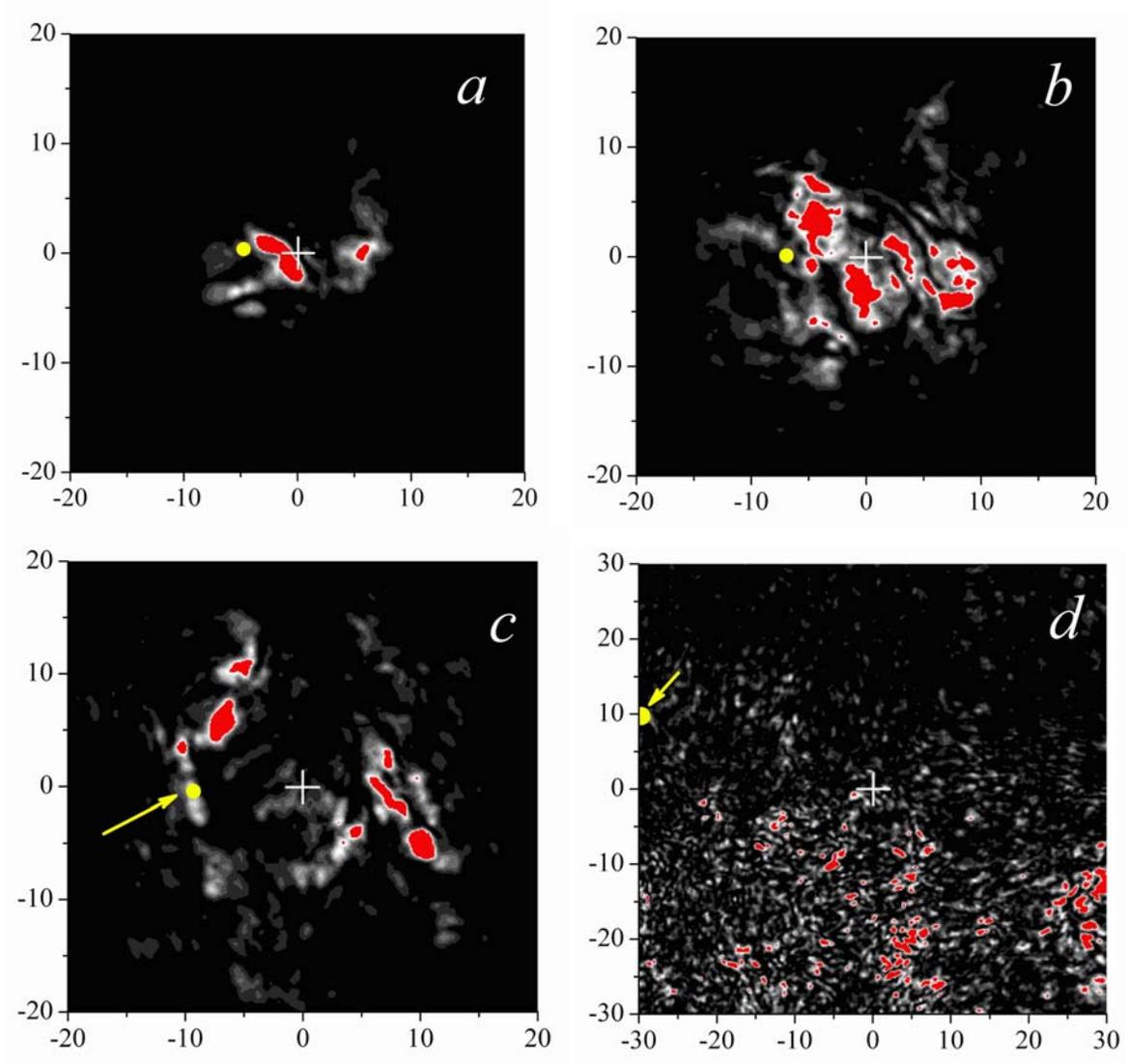

**Figure 26.** Beam intensity, $\tilde{I}(x, y)$, for those distances, $z$, corresponding to the points, $a - d$, in Fig. 25. In each case, a region in the $xy$-plane is shown in the system of coordinates whose origin is the center of the beam. Red indicates $\tilde{I}(x, y) \geq 0.2$. Yellow indicates the origin of the $z$-axis relative to the beam center (indicated by the white cross).

Due to these two factors, the average intensity, $I_{av}(r_w, z = D)$, (see Fig. 27b) essentially does not change in the region, $0 \le r_w \le r_{w,\max} \approx 32 cm$ (whereas for $C_n^2 = 2.5 \times 10^{-14} m^{-2/3}$, $I_{av}(r_w = 0, z = D)/I_{av}(r_w = r_{w,ma[}, z = D) \approx 3$. (See Fig. 5b.) Thus, the SI, $\sigma^2(z)$, at long distances should decrease noticeably with increasing turbulence strength, $C_n^2$, in contrast to the statement made in the paragraph 1 above for small propagation distances.

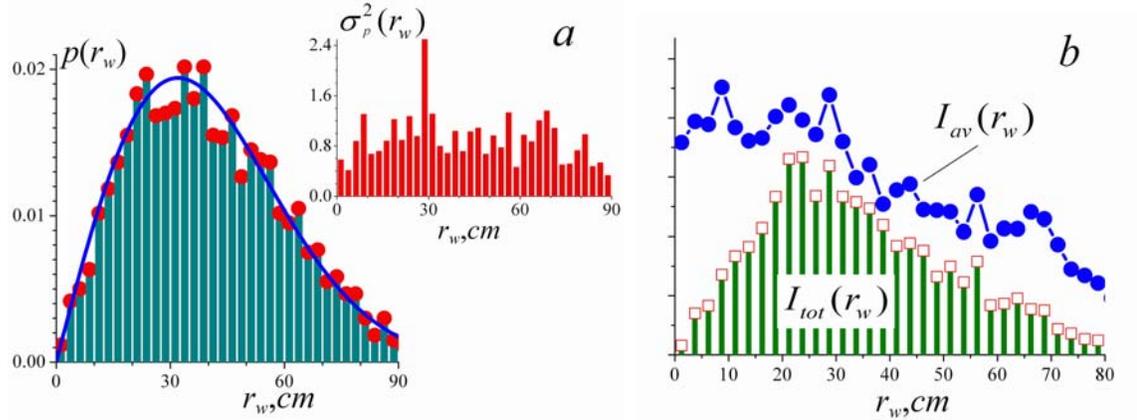

**Figure 27.** $a$ – The distribution function, $p(r_w)$, for a coherent beam propagating initially along the $z$-axis, at $z = D = 10 km$: $p(r_w) \approx 10^{-3} r_w \exp\left[-r_w^2/(2r_{w,\max}^2)\right]$, $r_{w,\max}^2 = 1025 cm^2$. $b$ – Statistical characteristics of intensity at the detector ($S = 1 cm^2$, $z = D$) as a function of the beam wandering, $r_w$; ● – $I_{av}(r_w)$ - average intensity produced by the beams shifted by the distance, $r_w$, (averaged over atmospheric states, $l$); □ – $I_{tot}(r_w)$ - combined contribution of all beams shifted by the distance, $r_w$. $I_{tot}(r_w) \sim I_{av}(r_w) \times p(r_w)$.

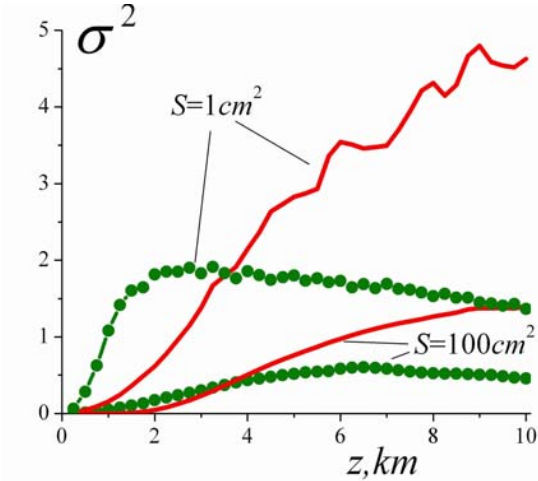

**Figure 28**. Comparison of SIs, calculated for different turbulent strengths. Red curves - $C_n^2 = 2.5 \times 10^{-14} m^{-2/3}$ (case A), green curves - $C_n^2 = 10^{-13} m^{-2/3}$ (case B).

The data in Fig. 28 confirm our prediction that $\sigma_B^2(z) > \sigma_A^2(z)$ for $z \le 4 km$ and $\sigma_B^2(z) < \sigma_A^2(z)$ for $z \ge 4 km$.

For the detector with the smaller area ($S = 1 cm^2$), the dependence of SI on distance, $z$, in the region where $\dfrac{\partial \sigma_{I(x,y)}^2}{\partial z} < 0$, can be described well enough by the empirical formula:

$$\sigma^2(z) \sim \sigma_{I(x,y)}^2(z) \cdot \langle r_w^2 \rangle^{1/4}. \qquad (22)$$

In agreement with (22), for $C_n^2 = 2.5 \times 10^{-14} m^{-2/3}$ and for $6km \leq z \leq 10km$ the increase of $\sigma^2(z)$ with distance is associated only with increasing $\langle r_w(z) \rangle$, because $\sigma^2_{I(x,y)}(z \geq 6km)$ decreases. (See Fig. 23, curve A.) In the more turbulent atmosphere, $C_n^2 = 10^{-13} m^{-2/3}$, the role of wandering is less pronounced due to the significant decrease in the level of fragmentation of the beam, $\sigma^2_{I(x,y)}(z)$, in the process of its propagation. (See Fig. 23, curve B at $z \geq 5km$.) The SI $\sigma^2(z > 5km; S = 1cm^2)$ decreases with distance even though $\langle r_w(z) \rangle$ increases.

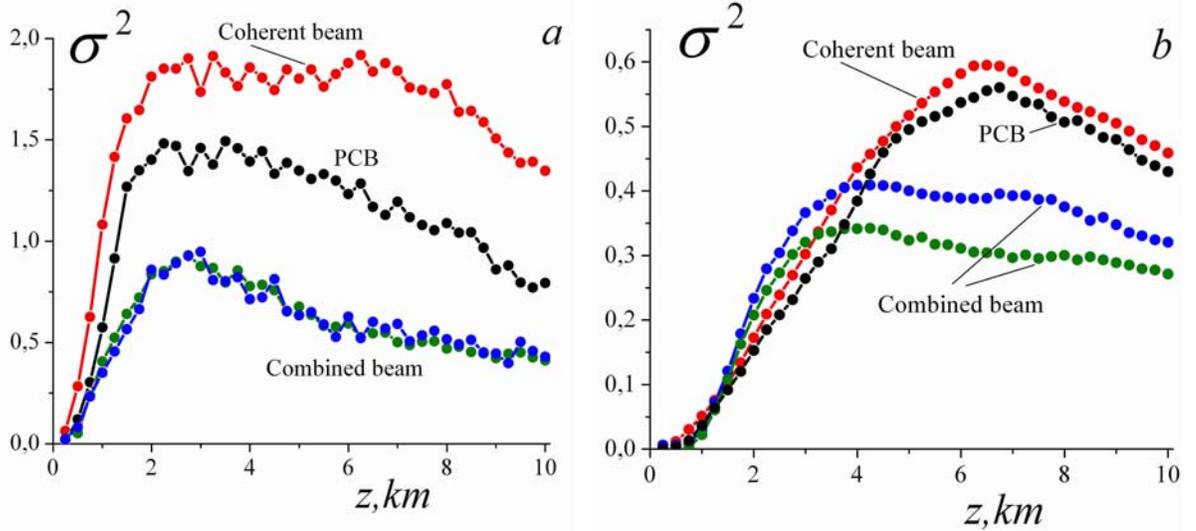

**Figure 29.** The decrease of SI in a strong turbulent atmosphere, $C_n^2 = 10^{-13} m^{-2/3}$, for different methods of generation of PCBs. a- $S = 1cm^2$, b- $S = 100cm^2$. $M = 6$. Red - $\sigma^2(z)$ for coherent beam; black – $\hat{\sigma}^2(z)$ for a PCBs produced by the phase mask, $\varphi_m(x,y)$, described in Sec. 2. (See Fig. 6). Green and blue – a combined beam was used (see Fig. 18) with the mask, $\varphi_0(r) = \chi \cdot r / r_{v0}$, $\chi = 4$ (blue) and $\chi = 6$ (green).

Note that for the detector with the larger area ($S = 100cm^2$) the role of wandering is more important. This statement is easy to prove, even though it is paradoxical at first glance. As one can see from Fig. 28, for $C_n^2 = 10^{-13} m^{-2/3}$ the maximum of $\sigma^2(z; S = 100cm^2)$ is noticeably shifted to the right relative to the maximum value of $\sigma^2(z; S = 1cm^2)$. The reason for this is that in the intensity field with low fragmentation, the dependence of the average signal at the detector, $I_{av}(r_w)$, on the parameter, $r_w$, increases with increasing detector area [*]. In other words, the role of beam wandering increases. The scintillation index, $\sigma^2(z)$, decreases only under conditions of significant spreading of the beam and weak fragmentation. In our case, these conditions are satisfied for $z > 7km$. (See Fig. 28.)

---

[*] If the detector area is small, then in the small-scale random speckle field (see, for example, Fig. 26d), the average (over realizations of the spackle-field) level of the signal at the detector weakly depends on the distance from the center of the beam. (See Fig. 27b.) A detector with a larger area averages the signal over a set of neighboring speckles, and this detector is able to register decreasing integral intensity as one moves away from the center of the beam.

Note that in the stronger turbulence atmosphere the problem of SI reduction becomes significantly complicated. The effectiveness of averaging over the states of the PCB, $m$, for a given atmospheric state, $l$, requires the statistical independence of signals at the detector, $I_m^l$ ($m = 1, 2, ..., M$). But just this condition is difficult to achieve, because at large $C_n^2$, the dependence of the values, $I_m^l$, on the initial state of the PCB, $U(x, y, z = 0)$, becomes weak.

The results of application of different types of PCBs are presented in Fig. 29. The use of a phase mask, $\varphi_m(x, y)$, that practically does not change the initial direction of the beam, in the case of detectors with small size (Fig. 29a), led to a reduction of SI (opposite to the case of relatively weak turbulence – see Sec. 2). However, this effect is not strongly related to the averaging over states of the PCB: $\hat{\sigma}^2(z = D) = 0.85$ - only by ~25% below the SI, $\sigma_s^2(z = D) = 1.1$, calculated for the individual pulses, $I_m^l$ ($m = 1, 2, ..., 6; l = 1, 2, ..., 300$). Namely, the result $\sigma_s^2(z = D) < \sigma^2(z = D)$ for $M = 1$ is related to the modification of the structure of the beam. As discussed above (in Sec. 2), the additional phase perturbations produced by the mask, increased the perturbation of the beam produced by the atmosphere. (See Fig. 10a, curve 3.) In the strong turbulence atmosphere, the opposite relation is realized. In this case, scattering of the beam by the phase mask results in a reduction of characteristic scales of the inhomogeneous speckle field at large distances and a reduction of $\sigma_s^2(z)$. But statistically independent states of the phase modulator, $\varphi_m(x, y)$, at given $l$ corresponds to strongly correlated values, $I_m^l$. The effect of averaging (decrease of $\hat{\sigma}^2(z)$ relatively to $\sigma_s^2(z)$) practically does not work even for $S = 1cm^2$. For the larger detector (Fig. 29b), the effect of using a PCB does not reveal itself: $\hat{\sigma}^2(z) \approx \sigma_s^2(z) \approx \sigma^2(z)$.

Application of a combined beam, with a time varying direction of propagation is more effective. (See Fig. 29a.) At the same time, we would like to mention an interesting fact. Increasing the deviation of the beam from the $z$-axis by increasing $\chi$ (which would have led to increase of statistical independence of values, $I_m^l$) did not influence the value of SI – the green and blue curves coincide within numerical errors. We observe saturation effects for the parameter, $\chi$. The calculated coefficient of effectiveness is, $\eta \approx 0.5$, in these cases. Namely, it was enough to use $M = 3$ (instead of $M = 6$) to derive a similar result for $\hat{\sigma}^2(z)$.

In conclusion, we discuss the use of large-size detectors (Fig. 29b). It is evident that for high-rate optical communication, detectors should be used with areas of $S \approx 100cm^2$ or greater. In this case, the reduction of $\hat{\sigma}^2(z)$ relative to $\sigma^2(z)$ is more complicated because when the area, $S$, of the detector increases, the correlations among $I_m^l$ increase (the other parameters of the system being held constant). An increase of $\chi$ leads to a decrease of $\hat{\sigma}^2(z)$ (compare the blue and green curves), but the final effect on the ratio, $[\sigma^2(z)/\hat{\sigma}^2(z)]$, is rather small. The parameter, $\eta \approx 1/3$ (green curve) indicates that it was sufficient to average over only two states of the combined beam instead of $M = 6$. But in this case, the small value, $\eta \approx 1/3$, should be taken with some optimism because this creates opportunities for significant improvement of the results using some special time-independent phase modulation, $\varphi_{0,opt}(x,y)$, that is asymmetric in azimuth $\beta$. Since the result presented in Fig. 29b for smooth phase mask, $\varphi_0(r) = \chi \cdot r / r_{v0}$, could be derived even by averaging over two positions of the maximum of combined beam, then the effective distribution, $\varphi_{0,opt}(x,y)$, is quite realistic. To make effective all six states of the combined beam, the inhomogeneities in $\varphi_{0,opt}(x,y)$ with characteristic size $\Delta\beta \sim 60^0$ have to be designed, which are large-scale enough to produce this result and decrease $\hat{\sigma}^2(z=10km)$ to the level $\sim 0.1$ (small-scale modulation were not effective, see black curves in Fig. 29).

**Conclusion**

In conclusion we summarize our results, which are quite useful for designing high-rate and long-distance optical communications through turbulent atmospheres:

- Significant suppression of the scintillation index (SI) at the detector can be achieved by averaging over a set of states, $m = 1, 2, ..., M$, of a partially coherent beam (PCB). However, the necessary statistical independence of the signals, $I_m^l$, for a given atmospheric state, $l$, required for the method to be effective, is still not guaranteed. Indeed, the turbulent atmosphere traversed by a set of PCBs could produce correlations among the intensities, $\{I_m^l\}_{m=1,2,...,M}$, which decrease the effectiveness of the averaging. Our numerical experiments show that the correlations increase with increasing turbulence. For a high level of turbulence, the SI can be relatively small due to forming a small-scale random speckle field. However, additional SI suppression using this averaging method is complicated.

- When using a phase mask with the random phase modulation, $\varphi_m(x,y)$, beam wandering is one of the main sources of correlations among the $\{I_m^l\}_{m=1,2,...,M}$ because of similarity in the trajectories of the beam center for different states of the PCB, $m$. Consequently, a phase modulator (that does not significantly change the initial direction of the beam) becomes ineffective.

- To reduce the SI, one must (i) compensate for beam wandering and (ii) maintain statistical independence of the signals, $I_m^l$. These requirements can be achieved by using a set of PCBs with preselected angles of propagation relative to the direction to the detector. The phase modulations, $\varphi_m(\vec{r}, z=0) = \vec{a}_m \cdot \vec{r}$, can be used for a predetermined $\{\vec{a}_m\}$ defined in terms of the statistical parameter, $\sqrt{\langle r_w^2 \rangle}$, that characterizes the beam wandering. The use of averaging in this case can produce significant reduction of the SI. However, the required high-frequency ($\sim 10^9 s^{-1}$) phase modulator is still a major obstacle.

- To overcome the limitations described above, a combined beam (Gaussian and an optical vortex) can be designed to compensate for wandering and eliminate the need for a high-frequency phase modulator. Designing the corresponding optimal optical system will require additional research to determine the best amplitudes and radii for the Gaussian beam and the optical vortex. An important remaining problem is determining the optimal time-independent phase modulation mask, $\varphi_{0,opt}(x,y)$.

**Acknowledgement**
This work was carried out under the auspices of the National Nuclear Security Administration of the U.S. Department of Energy at Los Alamos National Laboratory under Contract No. DE-AC52-06NA25396. This research was funded by the Office of Naval Research.